\documentclass[9pt]{article}
\usepackage{hyperref}%
\usepackage{thmtools}
\usepackage{thm-restate}
\usepackage{bbm}
\usepackage{algorithm,algorithmic}
\usepackage{amsmath}%
\usepackage{amssymb}%
\usepackage{adjustbox}
\usepackage{amsthm}%
\usepackage[round]{natbib}
\usepackage{amsmath,amssymb,amsthm,bm,enumerate,mathrsfs,mathtools}
\usepackage{latexsym,color,verbatim,multirow,longtable}
\usepackage{graphicx}
\usepackage{caption}
\usepackage{subcaption}
\usepackage{enumitem}
\usepackage{color}
\usepackage[margin=.7 in]{geometry}

\newtheorem{theorem}{Theorem}[section]

\newtheorem{assumptions}[theorem]{Assumption}

\newtheorem{lemma}[theorem]{Lemma}
\newtheorem{corollary}[theorem]{Corollary}
\newtheorem{remark}[theorem]{Remark}
\theoremstyle{definition}

\allowdisplaybreaks
\begin{document}
\fontsize{12}{14pt plus.8pt minus .6pt}\selectfont \vspace{0.8pc}
\centerline{\large\bf Detecting strong signals in gene perturbation experiments:}
\centerline{\large\bf An adaptive approach with power guarantee and FDR control}
\vspace{.4cm} \centerline{Leying Guan\textsuperscript{1}, Xi Chen\textsuperscript{1}, Wing Hung Wong\textsuperscript{1,2}} \vspace{.4cm} \centerline{\it
Departments of Statistics\textsuperscript{1} and Biomedical Data Sciences\textsuperscript{2}, Stanford University} \vspace{.55cm} \fontsize{9}{11.5pt plus.8pt minus
.6pt}\selectfont
{\large
 \begin{abstract}
The perturbation of a transcription factor should affect the expression levels of its direct targets. However, not all genes showing changes in expression are direct targets. To increase the chance of detecting direct targets, we propose a modified two-group model where the null group corresponds to genes which are not direct targets, but can have small non-zero effects. We model the behaviour of genes from the null set by a Gaussian distribution with unknown variance $\tau^2$, and we discuss and compare three methods which adaptively estimate $\tau^2$ from the data: the iterated empirical Bayes estimator, the truncated MLE  and the central moment matching estimator. We conduct a detailed analysis of the properties of the iterated EB estimate which has the best performance in the simulations. In particular, we provide theoretical guarantee of its good performance under mild conditions.

We provide simulations comparing the new modeling approach with existing methods, and the new approach shows more stable and better performance under different situations. We also apply it to a real data set from gene knock-down experiments and obtained better results compared with the original two-group model testing for non-zero effects.
\end{abstract}
\section{Introduction}

The transcriptional regulatory networks, formed by transcription factors(TFs) and their targets, are believed to play an important role regulating embryonic stem(ES) cell pluripotency(\cite{niwa1998self,niwa2000quantitative, chambers2004self, loh2006oct4, kim2008extended, chen2008integration}).  A multitude of inference methods exist in the literature for the identification of such networks using observational gene expression data(\cite{friedman2000using,murphy1999modelling, kim2004dynamic, lebre2010statistical}). On the other hand, there is also intense interest in using perturbation experiments in the study of gene regulation. For example, the TF knock-down experiment is expected to very informative  identifying potential targets of a TF because it depicts a less complex picture and can  provide evidence for causal relationships(\cite{geier2007reconstructing,werhli2006comparative}).

Traditionally, potential targets of the TF are usually identified as the subset of differentially expressed genes between the control and experiment group. However, when an important TF has been knocked down, it is almost always the case that the proportion of significantly changed genes is much larger than expected(\cite{ivanova2006dissecting, zhou2007gene}). As a concrete example, consider the data set analyzed in this study, which is from the knockdown experiment for two TFs which play an important role  regulating embryonic stem (ES) cell pluripotency(see section \ref{sec:real_data} for details). In this data set, the number of differentially expressed genes is very large, while the number of likely direct targets (from external CHIP-seq data assessing TF binding) is significantly smaller.

There are two popular explanations for this phenomenon:
\begin{enumerate}
\item The theoretic null distribution of the test statistics(often z-score and other analogous quantities) for zero effect is not accurate.
\item There are a large number of genes showing non-zero but small changes of gene expression level, as effects of the perturbation.
\end{enumerate}
Proposed solutions include modifying the null distribution of $z$-score empirically(\cite{efron2007correlation, efron2008microarrays}) and applying a cut-off to fold-change as a second-layer filter; the latter has been extremely popular in practice(\cite{nichols1998formation, zhou2007gene, vaes2014statistical}).  While both of these approaches can narrow down the selected, the former tackles the problem mainly based on the first explanation while the latter adopts the second  implicitly,  and results can be different in general(\cite{witten2007comparison}). For the knockdown experiment, the latter seems preferable because it considers both the change magnitude and the non-zero significant level, which is more related to what scientists care about; however, this approach lacks a natural quantitative justification.


Here, we propose a simple model to combine these two perspectives.  By using a Gaussian distribution with unknown variance to describe  the underlying behavior of genes in the null group, our model assumes that there can be relatively small non-zero effects even for the null genes. Assuming that the number of genes with large effect size is small, we test for the presence of such large effects relative to the background null variance.  Although this model is motivated by the knockdown experiment, it can be applied to more general multiple-testing setting where both the significance and the effect size matter. 

Our approach is related to the method of maximal agreement cut by  \cite{henderson2015making}. They similarly pointed out that testing approaches which measure evidence against the null hypothesis tend to over-populate the candidate list with those associated with small variance, while approaches that consider only the magnitude will overlook the noise. While sharing the same spirit, our method does not aim to find the top $\alpha\%$ subset of genes maximizing the expected overlap with the truth, with some assumed prior for all genes. Instead, we are interested in identifying the subset of genes which could not be described well by the prior describing the majority.

In the setting of knock-down experiment, our model describes a scenario different from the one assumed in approaches testing for differentially expressed genes. In our model, it is assumed that, perhaps due to propagation through the gene regulatory network, when we collect the data, a lot and even all genes may have been influenced once a TF has been knocked down, and we take this possibility into consideration.  We show that in this scenario, it is still possible to test  for strong effect if (1) the direct target tends to have larger effect size, and (2) there are enough null hypotheses to estimate the variance under the null.

The paper is organized as follows: We describe our model in section \ref{sec:model} and procedures to estimate the null variance in section \ref{sec:methods}. In section \ref{sec:proof}, we study the properties of one $\tau^2$ estimating procedure; in section \ref{sec:extension}, we extended the model to the non-centered case and the case of two sample testing with unequal variance. We provide simulations in section \ref{sec:sim} and real data examples in section \ref{sec:real_data}.

\section{Statistical model and estimation procedure}
\subsection{Statistical model }
\label{sec:model}
We assume there is a control group with $m_0$ replicates and an experiment group with $m_1$ replicates after knocking down one TF of interest. The expression levels for $N$ genes are measured for each replicate. Let $x_{i,j}$ be the measurement for gene $i$ in replicate $j$ from the experiment group, and $z_{i,j}$ be the measurement for gene $i$ in replicate $j$ from the control group. Without loss of generality, assume the mean level of $z_{i,j}$ is $0$ and the mean level of $x_{i,j}$ is $\mu_{i}$:
\begin{align*}
&x_{i,j} \sim N(\mu_{i}, \sigma^2_i),\;\; \forall i = 1,2,...,N, \;\; j = 1,2,...,m_1\\
&z_{i,j} \sim N(0, \sigma^2_i),\;\; \forall i = 1,2,...,N, \;\; j = 1,2,...,m_0\\
\end{align*}
To do inference on $\mu_i$, we can look at the two-sample test statistics, 
\begin{align*}
&\bar{x}_{i} - \bar{z}_{i} = \frac{\sum^{m_1}_{j=1}x_{i,j}}{m_1} - \frac{\sum^{m_0}_{j=1}z_{i,j}}{m_0}\sim N(\mu_i, \sigma^2_i(\frac{1}{m_1}+\frac{1}{m_0})),\;\;\forall i=1,2,...,N
\end{align*}
Or the paired sample test statistics to remove the batch effect ($m_0 = m_1$),
\begin{align*}
&\overline{x_i - z_i} = \frac{\sum^{m_1}_{j=1}(x_{i,j}-z_{i,j})}{m_1}\sim N(\mu_i, \sigma^2_i\frac{1}{m_1}),\;\;\forall i=1,2,...,N
\end{align*}

As there is no fundamental difference between these two tests in our later analysis,  we will omit the notation $z_{i}$ and use the following common notations for simplicity:
\begin{align*}
&\bar{x}_{i}\sim N(\mu_i, \sigma^2_{\bar{x}_i}), \forall i = 1,2,...,N\\
&\hat{\sigma}^2_{\bar{x}_i} \sim \sigma^2_{\bar{x}_i}\frac{\chi^2_{m-k}}{m-k}, \forall i = 1,2,...,N
\end{align*}
where  $ \sigma^2_{\bar{x}_i} = \frac{\sigma^2_i}{n}$, with $n$ being the effective sample size and $\hat{\sigma}^2_{\bar{x}_i}$ is the usual unbiased variance estimate of $\bar{x}_i$. In the two-sample case, $n = \frac{m_1m_0}{m}$, $m = m_1+m_2$, $k = 2$ and in the paired sample case,  $n = m=m_1$, $k = 1$.

Following the widely used two group model(\cite{efron2008microarrays}), let $A_0$, $A_1$ denote the sets of nulls and non-nulls respectively and  $\gamma = \frac{|A_1|}{N}$ denote the proportion of non-nulls. We assume that the $\mu_i$s in $A_0$ and $A_1$ are generated from different distributions:
\[
\mu_i \sim  \left\{\begin{array}{ll}N(0, \tau^2) & \forall i\in A_0\\
g_i(.) & \forall i\in A_1\end{array}\right.
\]

For each gene $i\in A_1$, $g_i$ is some unknown density function. The parameter $\tau$ can be viewed as describing the range of normal behaviour. 

In contrast to the original two-group model, which corresponds to $\tau = 0$, we allow $\tau$ to take positive values. By relaxing this assumption on $\tau$, we are able to detect relatively abnormal behavior compared with the background signal.  If we know $\tau$, the $p$-value for the new null hypothesis for gene $i$ can be derived. Let $u_i = \frac{\bar{x}_i}{\sqrt{\tau^2+\hat{\sigma}^2_{\bar{x}_i}}}$, $\bar{x}_i\sim N(0, \tau^2+\sigma^2_{\bar{x}_i})$, under the null hypothesis, $u_i$ is a Welch statistics(\cite{welch1947generalization}) in the limit case with the degree of freedom for the first ``variance estimate" $\tau^2$ being $\infty$. Usual analysis controlling False discovery rate (FDR) or family wise error rate (FWER) carries through under our extended model in this case.

We emphasize that the parameter $\tau$ itself is informative because it characterizes how influential a stimulus is -- in our case, how dramatically the whole system changes after we have knocked down a TF. The value of $\tau$ reflects the importance of the TF: it can be set according to either prior knowledge or estimated from the data. The second approach is usually more feasible, as we lack a quantitative characterization of this kind of importance, and it can vary under different environments even for the same TF.  
\subsection{Estimation procedure}
\label{sec:methods}
In this paper, we propose and compare three methods estimating $\tau^2$:  an iterated empirical Bayes estimate(ITEB) method, the truncated MLE method and the central matching(CM) method. The latter two methods have also been applied to estimating the empirical null distribution in the traditional two group test problem(\cite{efron2012large,efron2001empirical}), and we have adapted them to our problem here. ITEB is a new method that we develop for our problem.

 For $ i\in A_0$, $\bar{x}_i \sim N(0, \sigma^2_{\bar{x}_i}+\tau^2)$ marginalizing out $\mu_i$. If $A_0$ is known,  the empirical Bayes estimate of $\tau^2$ with estimated variance $\hat{\sigma}^2_{\bar{x}_i}$ for $\sigma^2_{\bar{x}_i}$ is given by $\hat{\tau}^2_{A_0} = \frac{1}{|A_0|}\sum_{i\in A_0}(\bar{x}^2_i-\hat{\sigma}^2_{\bar{x}_i})$. Let $\delta$ be a pre-determined small value, the adjusted form $\hat{\tau}^2_{A_0} = [\frac{1}{|A_0|}\sum_{i\in A_0}\bar{x}^2_i-(1+\delta)\hat{\sigma}^2_{\bar{x}_i}]_+$,  is often preferred to reduce the error for small $\tau^2$  and to ensure  non-negativity. Similar form of estimate is analyzed by \cite{johnstone2001chi,johnstone2001thresholding} in the context of estimating non-centrality of $\chi^2$-distribution with known variance.
 
Let $F_i(.)$ denote the distribution of $\frac{x^2_i}{\tau^2+\hat{\sigma}^2_{\bar{x}_i}} $ when $i\in A_0$, which is the distribution of the square of a Welch's statistic as mentioned before, and $\tilde{F}_i(x) = 1- F_i(x)$.  The ITEB procedure is given below, which starts from the whole set (as $A_0\cup A_1$) and then iteratively remove potential outliers on the tail based on the current estimate of $\tau^2$. It stops when no point needs to be further removed.

{\small
\vspace{-.2cm}
\noindent\rule{\textwidth}{0.4pt}\\
\vspace{-.4cm}\\
\centerline{\bf Iterated empirical Bayes estimation(ITEB) of $\tau^2$}

\noindent Input: $\{(\bar{x}_i, \sigma^2_{\bar{x}_i}), \forall i = 1,2,...,N\}$, significance level $\alpha_1$, $\alpha_2$ and $\delta$. By default, $\alpha_1 = 0.1$, $\alpha_2 = 0.01$  and $\delta = \sqrt{\frac{8}{N}}$.

\noindent Output: $\hat{\tau}^2$, the estimated $\tau^2$.

\noindent Initialization: $S_0 = \{1,2,\ldots, N\}$ be the initial estimate of the null set,  and $ \hat{\tau}^2_{S_0} =  \frac{[\sum^N_{i=1} \bar{x}^2_i - (1+\delta)\sum^N_{i=1} \hat{\sigma}^2_{\bar{x}_i}]_+}{N}$.

 \noindent For $k = 1,2,...$, do
\begin{enumerate}
\item Update the p value for each gene  $p_i =  \tilde{F}_i(\frac{\bar{x}^2_i}{\hat{\tau}^2_{S_{k-1}}+\hat{\sigma}^2_{\bar{x}_i}})$. The ordered p values from small to large are $p_{(1)}, p_{(2)}, \ldots, p_{(N)}$. Let $i^*$ be the largest index, such that  $p_{(i^*)} \leq \frac{i^*}{N}\alpha_1$.
\item Let  $J^1_{k} = \{i\in A: \;  p_i \leq p_{(i^*)}\}$, and  $J^2_{k} = \{i\in A: p_i \leq \alpha_2\}$ and remove $J_{k}:=J^1_{k}\cap J^2_{k}$.  Update  $S_k = S_{k-1}/J_{k}$ and $\hat{\tau}^2_{S_k} =  \frac{[\sum_{i\in S_k} \bar{x}^2_i - (1+\delta)\sum_{i\in S_k} \hat{\sigma}^2_{\bar{x}_i}]_+}{|S_k|}$.
\item If  $S_k = S_{k-1} $, return $\hat{\tau}^2 = \hat{\tau}^2_{S_k}$
\end{enumerate}

\noindent\rule{\textwidth}{0.4pt}\\
}

The detailed descriptions of the truncated MLE and the CM estimator are given in Appendix \ref{app:procedures}. In Appendix \ref{app:sim}, we compare  performances  for the three estimators in different scenarios and  discuss their strengths and weaknesses. ITEB was found to have better performance overall, especially when the  non-null proportion $\gamma$ is small. Thus, we will focus on ITEB and we provide detailed analysis of its properties.

\section{Properties of ITEB}
 \label{sec:proof} 
 We study the estimation quality of ITEB as the number of hypotheses $N\rightarrow \infty$.  For simplicity, we analyze the algorithm under following mild conditions and notations with $\delta = 0$. Let $\lambda(\alpha):= \max_i \tilde{F}^{-1}_i(\alpha)$.  The degree of freedom for the variance estimate is $m$ for all $i$, and $K :=$ the number of iterations needed for the algorithm to stop.  Since in this section we only use the mean level $\bar{x}_i$ and its estimated variance $\hat{\sigma}^2_{\bar{x}_i}$, with slight abuse of notation, let $x_i := \bar{x}_i$,  $\hat{\sigma}^2_{i} := \hat{\sigma}^2_{\bar{x}_i}$, $\sigma^2_i := \sigma^2_{\bar{x}_i}$. For the two levels $\alpha_1$ and $\alpha_2$ in the ITEB algorithm, we let $0<\alpha_1 < \frac{1}{2e}$ be a fixed value, and we let $\alpha_2\rightarrow 0$ at a slow rate to simplify the notations in the proof(we always let $\frac{N\alpha_2}{\log^2N}$ bounded away from 0).
\begin{assumptions}\label{ass:ass1}
The degree of freedom for variance estimates  $m\geq 5$ is a constant and the non-null proportion $\gamma < 1-c$ for some positive constant $c$. The ratio of variances of different genes is bounded: there exists a positive constant $C$ such that $\frac{\max_{i} \sigma^2_i}{\min_{i} \sigma^2_i} \leq C$.
\end{assumptions}
Without loss of generality, we rescale $\min_i \sigma^2_{i} = 1$, then $C = \max_i \sigma^2_i$. We do not require $\tau^2$ to be positive or a constant. It can be 0 or decay to 0 as $N\rightarrow \infty$. 
\begin{assumptions}\label{ass:ass2}
There exists constants $L$ and $\epsilon$, such that 
\[
\forall i\in A_1, \;\;E[x^2_i-(1+\epsilon)\hat{\sigma}^2_i  \;|\; x^2_i-(1+\epsilon)\hat{\sigma}^2_i \leq L(\tau^2+1)] \geq (1+\epsilon)\tau^2
\]
\end{assumptions}
\begin{remark}
If $i\in A_0$, we have $E[x^2_i-\hat{\sigma}^2_i] = \tau^2$.  Assumption \ref{ass:ass2} states that, for $i\in A_1$, $x^2_i-\hat{\sigma}^2_i$ has expectation non-negligibly bigger than $\tau^2$ and this condition is not purely driven by observations from its tail.
\end{remark}
\begin{assumptions}\label{ass:ass3}
The non-null proportion $\gamma\rightarrow 0$ as $N\rightarrow \infty$.
\end{assumptions}
\begin{theorem}(Lower bound of the variance estimate)
\label{thm:lower}
Let $R_K := |J_K\cap A_1|$, where $J_K$ is the rejected  set at the last step $k =K $ from ITEB. Let $\Delta_1 = \sqrt{\frac{\log N}{N}}(\tau^2+C)$, $t_l = \max(\frac{\log^2 N}{N}, \min(\frac{l}{N},2\alpha_2))$, $\Delta_{2,l}= 3(\tau^2+C)t_l\log \frac{1}{t_l}$ and $\tau^2_l=[\tau^2 - \Delta_1-\Delta_{2,l}]_+$.  Under Assumption \ref{ass:ass1} and  \ref{ass:ass2}, we have $P(\cup^{N\gamma}_{l=0}\{R_{K}= l, \hat{\tau}^2 \geq \tau^2_l\})\rightarrow 1$.
\end{theorem}

As $\frac{\Delta_1+\Delta_{2,l}}{\tau^2+C}\rightarrow 0$ for all $l$, Corollary \ref{corollary:lower} is a direct result of Theorem \ref{thm:lower}.
\begin{corollary}
\label{corollary:lower}
Under Assumption \ref{ass:ass1} and  \ref{ass:ass2}, for any $\delta > 0$,  $lim_{N\rightarrow \infty}P(\hat{\tau}^2 \geq [\tau^2-\delta(\tau^2+C)]_+)  = 1$.
\end{corollary}

\begin{theorem}(Upper bound of the variance estimate)
\label{thm:upper}
Under Assumption \ref{ass:ass1} and \ref{ass:ass3}, suppose $\alpha_1>0$ is fixed and  $\alpha_2\rightarrow 0$ at a rate slow enough: $\gamma \lambda(\alpha_2)\rightarrow 0$. 
Then, for any $\delta > 0$, we have $\lim_{N\rightarrow \infty} P(\hat{\tau}^2 \leq \tau^2+\delta(\tau^2+C)) = 1$.
\end{theorem}

We next show that these results can usually lead to good performance in the follow-up analysis in practice.  Theorem \ref{thm:FDR} states that our estimate of $\tau^2$ can successfully control the FDR if we reject the hypotheses in the set $J_K$.  
\begin{theorem}(FDR control)
\label{thm:FDR}
Under Assumption \ref{ass:ass1} and \ref{ass:ass2}, if we reject all hypothesis in $J_K$, we have $\lim_{N\rightarrow \infty }FDR \leq \alpha_1$. 
\end{theorem}
\begin{remark}
Note that at the given level $\alpha_1$, $\alpha_2$ in the ITEB algorithm, $J^1_K$ will correspond to the set of rejections using the BH(Benjamini-Hochberg procedure)(\cite{benjamini1995controlling}) and $J_K$ will correspond to the set of rejections which are both rejected by the BH procedure and with p-values no greater than $\alpha_2$. The extra requirement that the p-value is no greater than a reasonable small value $\alpha_2$ is desirable in a lot of large scale hypotheses testing setting, including the knock-down experiment. 
\end{remark}

\begin{theorem}(Power analysis)
\label{thm:optimality}
Let $\phi_{i, \alpha} = \mathbbm{1}_{p_i \leq \alpha}$ and let $\phi^*_{i, \alpha}$ be the oracle decision rule knowing $\tau^2$: for any level $\alpha$, $\phi^*_{i, \alpha} = \mathbbm{1}_{x^2_i> \tilde{F}^{-1}_i(\alpha)(\hat{\sigma}^2_i+\tau^2)}$.  Let  $z^2_i = \frac{x^2_i}{\tau^2_i+\sigma^2_i}$ where $\tau^2_i = E[\mu^2_i]$ for $i\in A_1$. Under Assumption \ref{ass:ass1} and \ref{ass:ass3}, if we further assume that the density of $z^2_i$ is upper bounded by a constant, and the tail probability for $z^2_i $ decays sufficiently fast: 
 \[
 \lim_{w\rightarrow \infty}\sup_{\delta > 0}\sup_{i\in A_1}\frac{P(z^2_i \leq w(1+\delta))}{P(z^2_i \leq w)(1+\delta) } \leq  1
 \]
 Then we have $ \lim_{N\rightarrow \infty}\sup_{i\in A_1}\sup_{\alpha\geq 0}(P(\phi_{i, \alpha} = 1 ) - P(\phi^*_{i, \alpha} = 1)) \geq 0$.
 \end{theorem}
\begin{remark}
Recall that the follow-up  p value $p_i$ for hypothesis $i$ is $\tilde{F}^{-1}_i(\frac{x^2_i}{{\hat{\tau}^2 + \hat{\sigma}^2_i}})$. Thus Theorem \ref{thm:optimality} says that the test $\phi_i$ based on the estimated variance $\hat{\tau}^2$ is asymptotically as powerful as the optimal test based on the (unknown) true variance $\tau^2$.
\end{remark}

Proofs of Theorem ~\ref{thm:lower}, ~\ref{thm:upper},  ~\ref{thm:FDR} and ~\ref{thm:optimality} are given in the Appendix ~\ref{app:theorem}.
\section{Extension to two sample test with unequal variance}
\label{sec:extension}
For hypothesis $i$, the observations from the experiment and control groups, $x_{i,j}$ and $z_{i,j}$, can have different variances.  It is straightforward to generalize ITEB to this situation if we want to perform a two-sample test. We know that ITEB takes in $\{\bar{x}_i - \bar{z}_i\}$ and  $\{\hat{\sigma}^2_{\bar{x}_i - \bar{z}_i }\}$. In the unequal variance setting, we can estimate  $\hat{\sigma}^2_{\bar{x}_i - \bar{z}_i }$ by:
\[
\sigma^2_{\bar{x}_i- \bar{z}_i} = \frac{\sum^{m_1}_{j=1}(x_{i,j} - \bar{x}_i)^2}{m_1(m_1-1)}+ \frac{\sum^{m_0}_{j=1}(z_{i,j} - \bar{z}_i)^2}{m_0(m_0-1)}
\]
The degree of the $\hat{\sigma}^2_{\bar{x}_i- \bar{z}_i}$ approximated by
\[
df_{\sigma} = \frac{(\frac{\sum^{m_1}_{j=1}(x_{i,j} - \bar{x}_i)^2}{m_1}+\frac{\sum^{m_0}_{j=1}(z_{i,j} - \bar{z}_i)^2}{m_0})^2}{\frac{1}{(m_1 - 1)}(\frac{\sum^{m_1}_{j=1}(x_{i,j} - \bar{x}_i)^2}{m_1})^2+\frac{1}{(m_0-1)}(\frac{\sum^{m_0}_{j=1}(z_{i,j} - \bar{z}_i)^2}{m_0})^2}
\]
We approximate $F_i(.)$,the distribution of  the test statistics $\frac{(\bar{x}_i - \bar{z}_i)^2}{\tau^2+\sigma^2_{\bar{x}_i- \bar{z}_i} }$, by $F_{1, df}(.)$, the $F$ distribution with degree of freedoms $(1, df)$(\cite{satterthwaite1946approximate}), where $df$ is approximated by $df =(\frac{\tau^2}{\sigma^2_{\bar{x}_i- \bar{z}_i}}+1)^2 df_{\sigma}$.

\section{Simulation: Detection of  large signal}
\label{sec:sim}
We consider the  two-sample setting with equal-variance and generate data under various values of $\tau$ and non-null proportion $\gamma= \frac{|A_1|}{N}$. Specifically, we fix $N = 15000$, $m = 5$ for both the 
control and experiment group,  for any given $\tau$ and $\gamma$, where $\gamma = 1\%, 5\%$ and $\tau = 0, 0.1,...,1, 1.5, 2, 2.5,3$, we generate the true mean and variance  as below.
\begin{enumerate}
\item  Let $\mu_i = 0$ in the control group, and in the experiment group, we generate them as following:
 $$\mu_i \sim  \left\{\begin{array}{ll}N(0, \tau^2)&\forall i\in A_0\\
\pm U[1, \max(3, 10\tau)]&\forall i\in A_1\\
\end{array} \right.$$

 where $U[1,\max(3, 10\tau)]$ is the uniform distribution between $1$ and $\max(3, 10\tau)$, and the signs of $\mu_i$s will be half positive and half negative.

\item  We sample the variances $\sigma^2_i$ from its empirical distribution from the real data set, and we scale them to have mean level 1. 
\end{enumerate}
We compare the following approaches:
\begin{itemize}
\item ITEB estimate of $\tau^2$ and followed the Welch's t-test as previously described.
\item t-test for the null hypothesis testing for zero effect
\item EBarray(two group empirical Bayes method with each gene being either differential expressed or not),(\cite{kendziorski2003parametric,yuan2006unified}), and we choose the ``LLN" model to fit the data. Different genes are considered to have a common variance in the null group(or in the non-null group). 

\item EBarray, and we choose the ``LLNMV" method to fit the data. Different genes are considered to have difference variances with the inverse chi-square prior.
\end{itemize}

Figures (\ref{fig:sim_signal1}) and (\ref{fig:sim_signal2}) provide ROC curves(average sensitivity versus FDP) across 20 repetitions as the significant level in the testing step is varied.  We see that the new approach performs the best across different experiments. When $\tau = 0$, ITEB procedure and t-test behave similarly and have the best performance, and the same thing happens for ITEB and EBarray with ``LLN" method when $\tau$ is large. ITEB and EBarray with ``LLNMV" method are both stable across a wide range of $\tau$, but the ITEB procedure is more powerful in our simulations.

\begin{figure}
\begin{center}
\includegraphics[width = 1\textwidth, height = 1\textwidth]{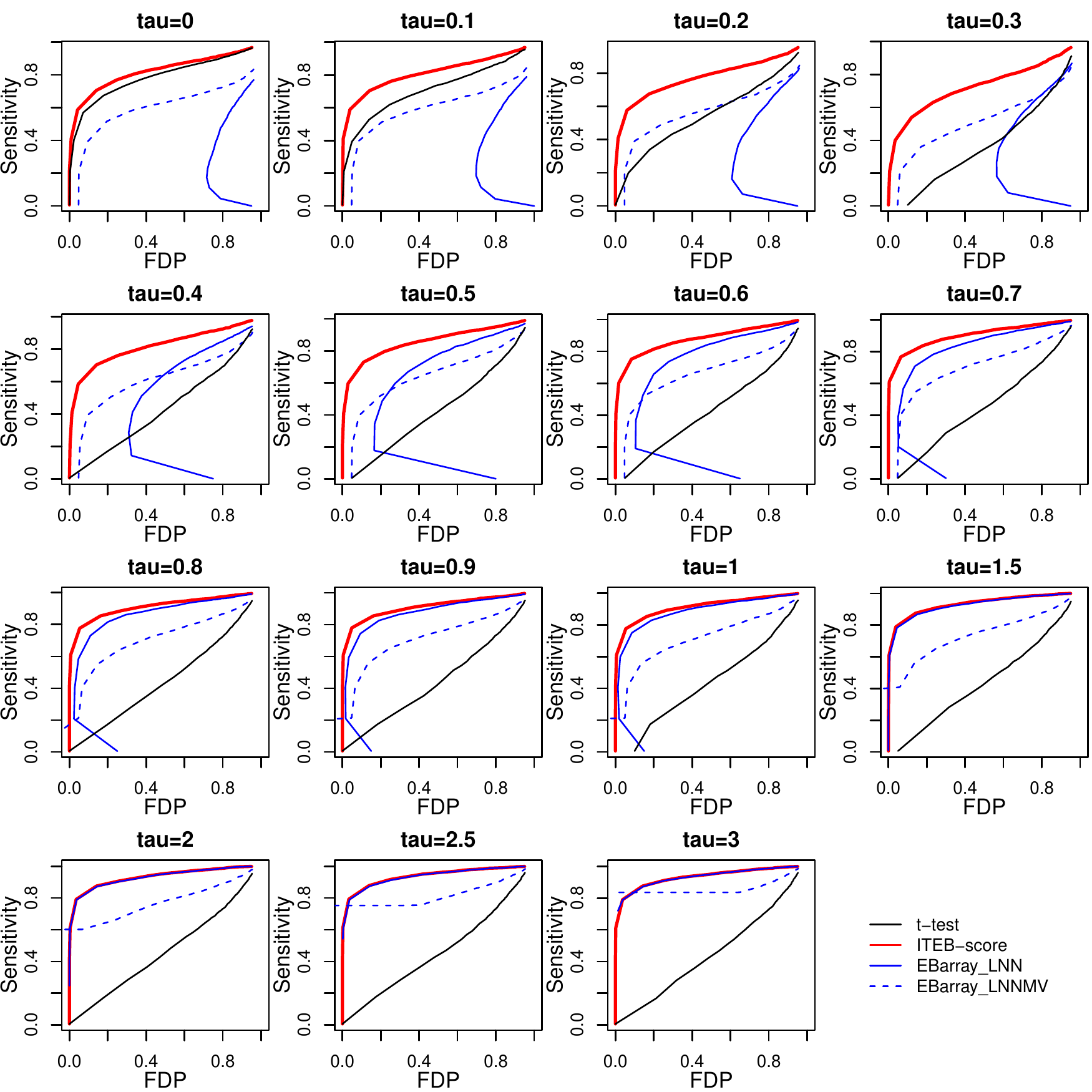}
\end{center}
\caption{\em ROC curve for $p = 1\%$.}\label{fig:simulationSignal_c1}\label{fig:sim_signal1}
\end{figure}

\begin{figure}
\begin{center}
\includegraphics[width = 1\textwidth, height = 1\textwidth]{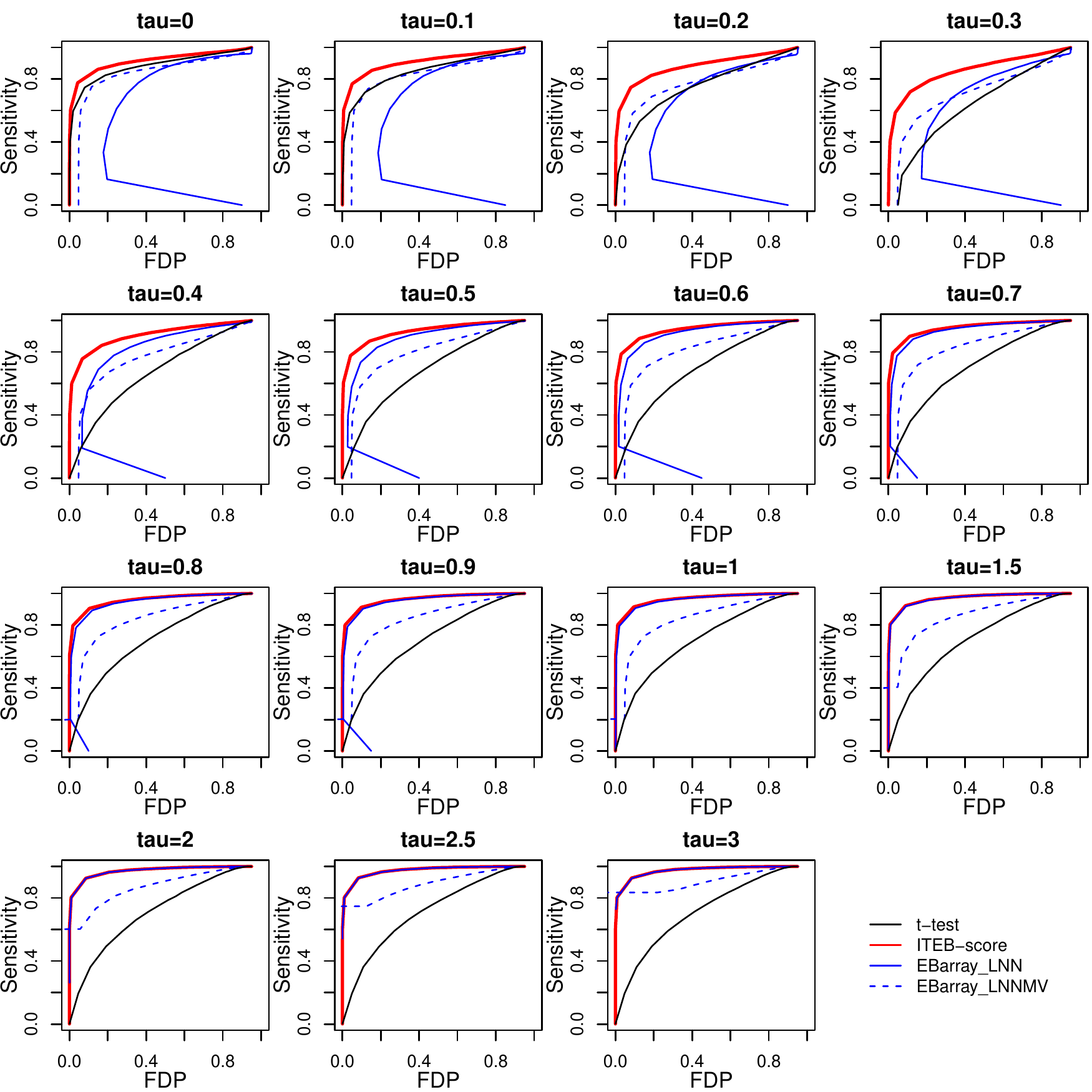}
\end{center}
\caption{\em  ROC curve for  $p = 5\%$.}\label{fig:simulationSignal_c2}\label{fig:sim_signal2}
\end{figure}

\section{Real Data Examples}\label{sec:real_data}
In this section, we apply our approach to data from two knock-down experiments described below. The quality of the results are evaluated by the enrichment of ChIP-seq peaks (for the perturbed TF) in active enhancers/promoters for the selected genes. Note that the ChIP-seq data for ES cells are external to the data used to select the genes, and it provides an orthogonal information to access how likely the selected genes are direct targets of the TFs. 

We perform gene knock-down experiments on 2 TFs on the mouse ES cell line R1. For each TF, RNA interference (RNAi) delivered using nucleofection was used to knock down its expression.  Puromycin selection was introduced 18 h later at 1 $\mu$ g/ml, and the medium was changed daily. 30 h, 48 h, and 72 h after puromycin selection, the cells were collected for RNA isolation.  After the experiments, Microarray hybridizations were performed on the MouseRef-8 v2.0 expression beadchip arrays (Illumina, CA).  More details of the experiments can be found in Appendix \ref{app:experiment}. Quantile normalization is performed in the first step to reduce the batch effect, and for the same reason, for each sample in the experiment group, we consider the paired test statistics with each pair being a pair of independent experiment and control samples from the same batch and time point. We have 8 paired observations for both POU5F1 and NANOG, and we take the log difference between the gene expression levels in a knock-down sample and its corresponding control sample  to further reduce the batch effect.  Table \ref{tab:data_info} summarizes the data we have at different time points.  Figure \ref{fig:tsne} shows results from nine realizations of the t-SNE(\cite{maaten2008visualizing}) plot using the top $1000$ genes with largest variance across experiments.  Each data point in the t-SNE plot  represents one sample (paired) in the experiment. We use the colors black, red and green to represent data at time points 30 hr, 48 hr and 72 hr respectively.  From the results, we see that differences between time points within the same knock-down experiment is comparable to the differences across batches, and they are very small compared with the differences across knock-downs . To compare the targets of two TFs, we will regard the different times points in the same knock-down as replicates of each other.
\begin{figure}
\begin{center}
\includegraphics[width =.8\textwidth, height = .8\textwidth]{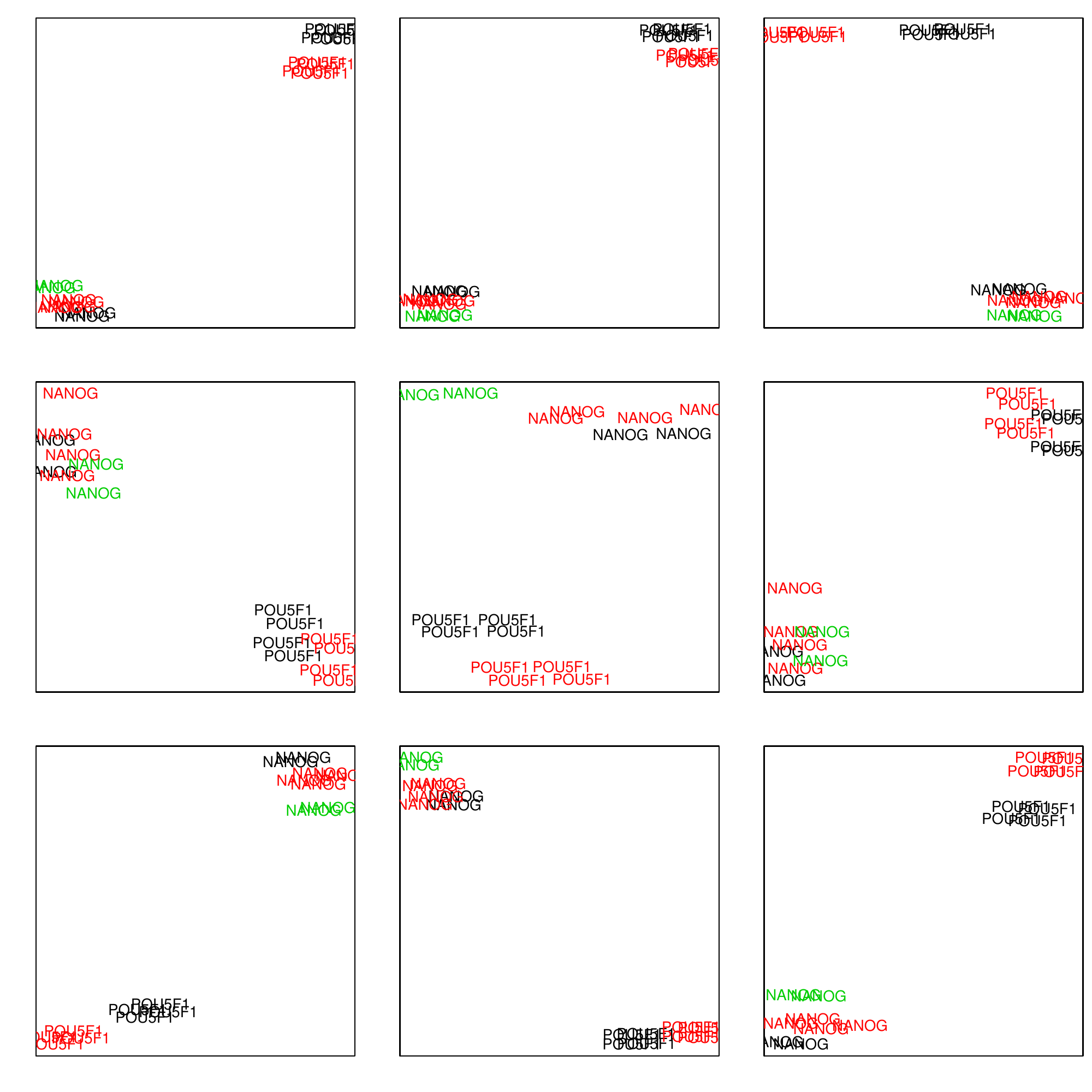}
\caption{\em Clustering with t-SNE algorithm. Black, red, and green colors represent experiments from 30 hr, 48 hr and 72 hr respectively. Note that there are two replicates from the same batch in a single experiment (same day, same TF, same batch) that are almost identical to each other.}\label{fig:tsne}
\end{center}
\end{figure}


\begin{table}
\centering
\caption{Information of  the knock-down data sets. }\label{tab:data_info}
\begin{tabular}{|l| l| l| l| l| }
\hline
& 30 hr&  48 hr &  72 hr\\\hline
POU5F1&4 pairs&4 pairs&--\\\hline
NANOG&2 pairs &4 pairs&2 pairs\\\hline
\end{tabular}
\end{table}

\noindent\textbf{ChIP-seq data and enhancer-gene association data}: To evaluate the quality of the selected gene set, we utilize two external data sets: the ChIP-seq data is from \cite{chen2008integration} and the enhancer-gene association data is from \cite{mumbach2017enhancer}. The ChIP-seq data contains results using chromatin immunoprecipitation coupled with ultra-high-throughput DNA sequencing (ChIP-seq) to map the binding locations of 13 sequence-specific TFs, including POU5F1 and NANOG. The enhancer-gene association data is generated by using the HiChIP method where the authors performed H3K27ac HiChIP in mouse ES cells.  H3K27ac is a histone modification mark characteristic of  active enhancers and promoters in the cell. HiChIP using H3K27ac mark as bait will provide enhancer-gene interaction information.
We can evaluate the quality of the selected gene set by examining  whether the binding sites of POU5F1 and NANOG are enriched near the active enhancers/promoters of the selected genes in the ES cell. 

Let us call the approach based on p-value using  a simple t-test $S_0$, and the approach based on p-value using  ITEB  $S_2$. We will focus on those genes with significant decrease in their expression levels after POU5F1/NANOG knock-down.  For each method, we set the cut-off using the BH procedure with targeted FDR level at $0.01$.  Accordingly, we set the the cut-off level $\alpha_1 = 0.01$ and we select 87 genes after knocking down POU5F1 and 43 genes after knocking down NANOG using $S_2$. These numbers are 2274 and 1267 using $S_0$. $S_0$ does not provide informative candidate lists with this criterion. To have a meaningful comparison, we also consider the case where we control FWER at 0.01, which is quite a stringent criterion and under which, $S_0$ selects 144 genes for POU5F1 and 49 genes for NANOG.  

We say that  there is supporting evidence of a gene being a  direct target  of a TF if this TF has at least one ChIP-seq peak within $x$ kilobase(kb) away from the gene's active enhancers/promoter.  As we change $x$ in a large range of value, Figure \ref{fig:chipCurve} gives the percentages of genes with this supporting evidence in the selected gene sets using $S_0$ using FWER and $S_2$ using FDR.  Figure \ref{fig:chipCurve} also shows the percentages of all genes with this supporting evidence(referred to as ``all" in the figure), and the percentages of bottom 2000 genes with this supporting evidence(referred to as ``bottom"). The bottom 2000 genes are those showing the smallest change after knock-down experiment(we consider this set to be the negative control). From Figure \ref{fig:chipCurve}, we see that the ChIP-seq enrichment is quite significant for the selected genes comparing with both the negative controls and all genes. The selected gene set using $S_2$ is significantly better than the gene set selected using $S_0$. Table \ref{tab:FWER} shows the result with $x = 20$.

\begin{figure}
\begin{center}
\includegraphics[width = .7\textwidth]{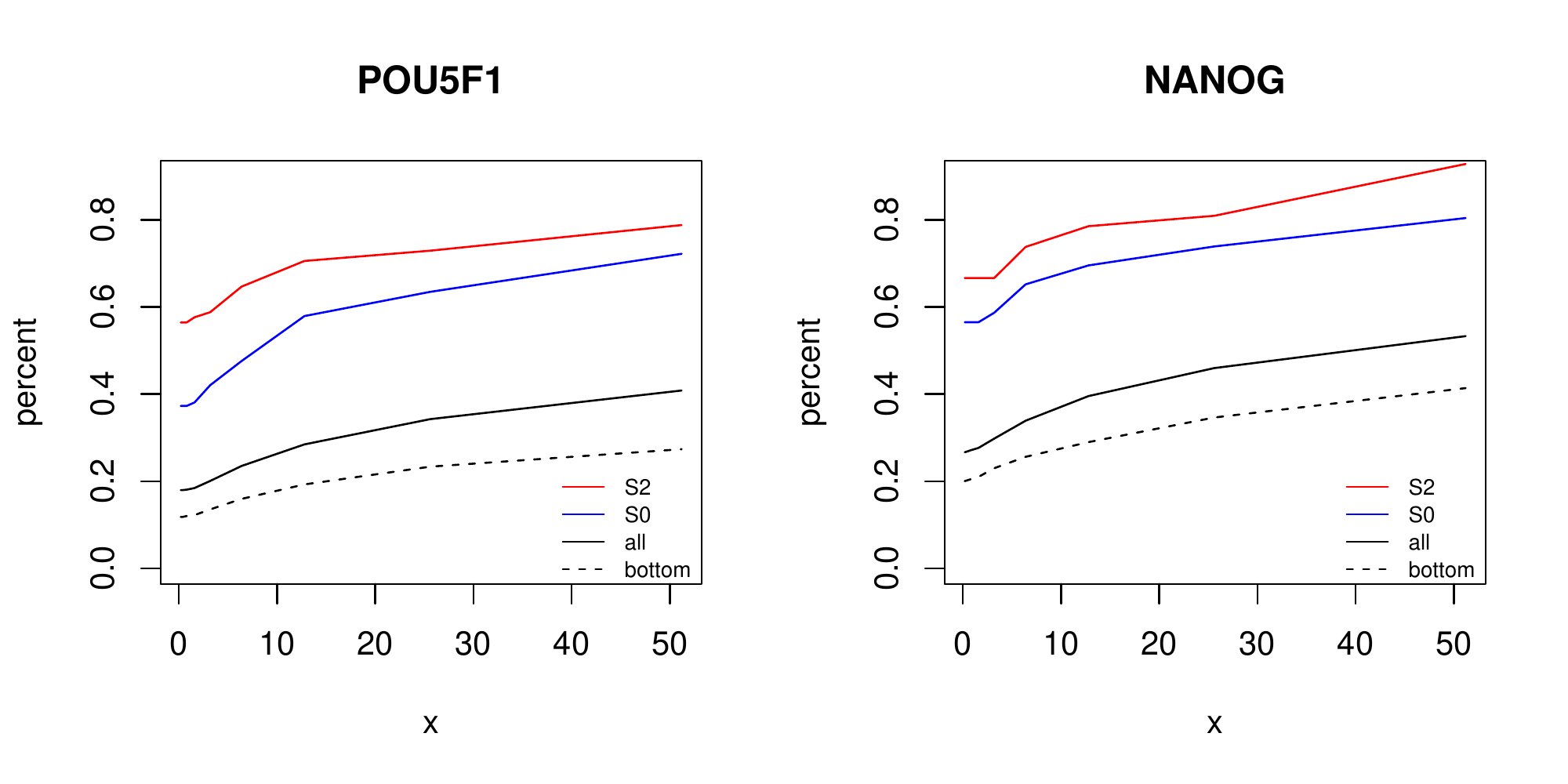}
\caption{\em Percent of genes with ChIP-seq nearby versus $x$ for the selected gene sets ($S_0$: genes selected using  t-test with FWER control, $S_2$: genes selected based on ITEB with FDR control, bottom: the bottom 2000 genes which show the smallest changes, all: all genes). The x-axis is the $\log(x)$, the threshold we use to define whether a gene has a ChIP-seq peak near its enhancer/promoter and the y-axis is the percentage of selected genes with ChIP-seq nearby.}\label{fig:chipCurve}
\end{center}
\end{figure}

 \begin{table}[ht]
\caption{$S_0 $, $S_2$ results with FDR/FWER level set at $0.01$}\label{tab:FWER}
\centering
\begin{adjustbox}{width = .7\textwidth}
\begin{tabular}{|l| l| l| l| l| l| l| }
  \hline
  &TF &percent(negative control)& size($S_0$ ) & percent$(S_0)$  & size($S_2$) & percent($S_2$)  \\ \hline
FDR& POU5F1 & 0.21 & 2274&0.48&87&0.73\\ \hline
  &NANOG & 0.32 &1267& 0.62&43&0.81 \\ \hline
  FWER& POU5F1 & 0.21 & 144&0.62&31&0.74 \\ \hline
&NANOG & 0.32 & 49& 0.74&20&0.8\\  \hline
\end{tabular}
\end{adjustbox}
\flushleft
\em{The first, third and fifth column are the percent of genes with Chip-seq+Hi-C support of the negative control set, genes selected by $S_0$ (blue, genes selected by  t-test) and  $S_2$ (red, genes selected based on ITEB) respectively.}
\end{table}

Not only $S_2$ provides a candidate gene set with much smaller size and higher quality, it provides a gene ranking list different from $S_0$. We fix $x = 20$ and consider the top $k$ genes using $S_0$ and $S_2$. Figure \ref{fig:geneCurve} shows that $S_2$ provides a ranking list with higher quality.
\begin{figure}
\begin{center}
\includegraphics[width = .7\textwidth]{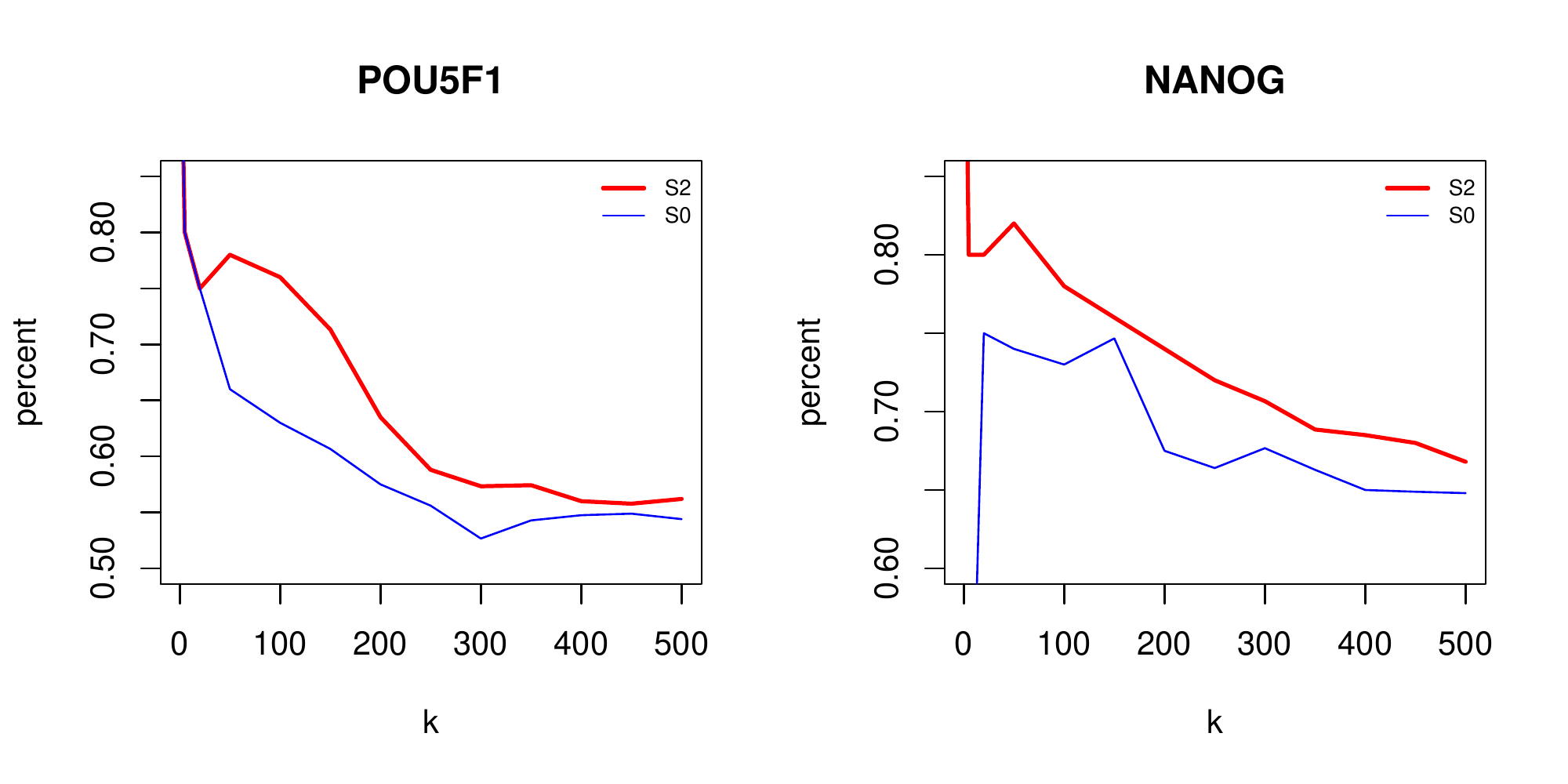}
\caption{\em Percent of genes with ChIP-seq nearby versus versus selected gene size. The x-axis is $k$, the threshold of the ranking on the whole list created using $S_0$ and $S_2$, and we only consider the top $k$ genes from each ranking list.}\label{fig:geneCurve}
\end{center}
\end{figure}

\section{Discussion}
\label{sec:discuss}
In this paper, motivated by the problem of identifying TF targets based on data from the knock-down experiment, we have proposed to test for large effect size instead of non-zero effect size in the two-group model where a Gaussian distribution with non-zero variance is used for the effect in the null group. We have considered three approaches(ITEB, truncated MLE and CM) to estimate this  non-zero variance adaptively,  and recommend ITBE for its computational efficiency, strong performance in simulation and attractive theoretical properties. Although we have focused on the Gaussian setting here, the idea of testing for strong signal and the approaches to estimate the null distribution can be applied to problems involving other data types.

The model  itself is related to the ``g-modeling"(\cite{carroll1988optimal, efron2014two}), the ``$\epsilon$-contamination"(\cite{huber1964robust, chen2016general}) and the ``robust Bayesian analysis"(\cite{berger1986robust, gaver1987robust, berger1994overview}). However, it should be noted that our approach  has a different goal from the g-modellings. We only estimate the shape of the null distribution of the mean parameter  while the g-modeling models the marginal distribution of the mean parameter considering both the nulls and the non-nulls.  Our purpose for estimating the null effect distribution is to set a cut-off for the strong signals adaptively, which is not the case for the g-modelling. Although our model can be considered as a special case of the $\epsilon$-contamination model in the parameter space and a special case of the robust empirical Bayesian analysis, the use of these models for large effects have not been studied and, to the best of our knowledge, methods with provable guarantees on power and FDR have not been demonstrated previously.

\vskip .2in
 {\bf Acknowledgments}
We would like to thank Zhou Fan and Professor Bradley Efron for the helpful discussions and suggestions. We would also like to thank Zhana Duran for organizing the enhancer-gene association data set.  This work is partially supported by NIH Grants R01HG007834 and R01GM109836.

\appendix
\section{Proof of Theorem \ref{thm:lower}, Theorem \ref{thm:upper},  Theorem \ref{thm:optimality} and Theorem \ref{thm:FDR}}
\label{app:theorem}
Let $\Delta_1 := \sqrt{\frac{\log N}{N}}(\tau^2+C)$, and let  $t_l := \max(\frac{\log^2 N}{N}, \min(2\alpha_2,\frac{l}{N})), \;\Delta_{2,l} := 3(\tau^2+C)t_l\log \frac{1}{t_l}, \;\tau^2_l := [\tau^2 -\Delta_1-\Delta_{2,l}]_+$  for all  $ l = 0, 1,\ldots, |A_1|$. Let the oracle estimator be defined as $\tau^2_* = \frac{\sum_{i\in A_0} x^2_i - \hat{\sigma}^2_i}{|A_0|}$. Let $p_{i, l} = \tilde{F}(\frac{x^2_i}{\tau^2_l +\hat{\sigma}^2_i})$ be the p-values calculated using  $\tau^2_l$ and let $p_{(i), l}$ be the ordered null p-values  from small to large.  Let $B_{1,l} =  \{i\in A_0:  p_{i,l} \leq p_{(s_0),l}\}$,  where $s_0 = \max \{s: p_{(s),l} \leq \frac{(s+l)\alpha_1}{N}\}$, and $B_{2,l} = \{i\in A_0: p_{i, l} \leq \alpha_2\}$.  Lemma \ref{lem:tailDecays}  contains the deterministic relationships we will use later.
\begin{lemma}
\label{lem:tailDecays}
Suppose $0<\alpha_1 < \frac{1}{2e}$ to be fixed and $\alpha_2\rightarrow 0$ at a slow rate ($\alpha_2\frac{N}{\log^2 N}$ is bounded away from 0). Under Assumption \ref{ass:ass1},  the following statements hold:

(1)$\lambda(\frac{l\alpha_1}{N}) \lesssim (\frac{N}{l})^{\frac{2}{5}},\;\frac{\Delta_1}{\tau^2+1}\lambda(\frac{\alpha_1}{N}) \rightarrow 0, \;\sup_{l\geq 1}\frac{\Delta_{2,l}}{\tau^2+1}\lambda(\frac{l\alpha_1}{N}) \rightarrow 0, \;\sup_{l\geq 0}\frac{\Delta_{2,l}}{\tau^2+1}\lambda(\alpha_2) \rightarrow 0$

 (2)$\lim_{N\rightarrow \infty}\sup_{i\in A_0}\sup_{l\geq 0}\sup_{\alpha \geq\min\left(\alpha_2, \frac{(l\vee 1)\alpha_1}{N}\right)}  \frac{P(p_{i,l} \leq \alpha)}{\alpha} = 1$
 
(3)The estimate $\hat{\tau}^2_{S_k}$ is non-increasing in the iteration number $k$ in the ITEB procedure.
\end{lemma}  
\noindent Proof of \ref{lem:tailDecays} is deferred to section \ref{app:lem}.
\subsection*{Proof of Theorem ~\ref{thm:lower}}
\begin{proof}
Let $E_{k, l} := \{\hat{\tau}^2_{S_k} \geq \tau^2_l\}$. Because $\cup^{|A_1|}_{l=0}\{R_K = l\}$ is a partition of the full space, to prove the statement, we show the following:
\[
P(\cup^{|A_1|}_{l = 0}\{R_K =  l, E^c_{K, l}\})\leq P(\cup^{|A_1|}_{l = 0}\{R_K \leq  l, E^c_{K, l}\}) \rightarrow 0
\]
Define $R_k := |J_k\cap A_0|$ for each iteration $k$. We also let $R_0 = 0$, and $R_{k-1}\leq R_{k}$ as a consequence of Lemma \ref{lem:tailDecays}, part (3). We prove it by showing that the events $\{R_{k} \leq  l, E^c_{k, l},  E_{k-1, l}\}$ and $\{E^c_{0, l}\}$ do not happen on a properly chosen event $M$ which itself happens with probability approaching 1.  The existence of such $M$ is sufficient for our argument:
\begin{align*}
\sum^{|A_1|}_{l = 0} P(\{R_K\leq l, E^c_{K,l}, M\})\leq &\sum^{|A_1|}_{l = 0} P(R_{K} \leq l, E^c_{K,l}, E_{K-1,l}, M)+\sum^{|A_1|}_{l= 0} P(R_{K} \leq l, E^c_{K-1,l}, M)  \\
 \leq& \sum^{N\gamma}_{l= 0} P(R_{K-1} \leq l, E^c_{K-1,l}, M)\leq \ldots \leq \sum^{|A_1|}_{l=0}P(E^c_{0,l}, M)  = 0
\end{align*}
Then $P(\cup^{|A_1|}_{l = 0}\{R_K \leq  l, E^c_{K, l}\})\leq P(\cup^{|A_1|}_{l = 0} \{R_K\leq l, E^c_{K,l}, M\}) +P(M^c)\rightarrow 0$.
We now find $M$ which contradicts $\{R_{k} \leq  l, E^c_{k, l},  E_{k-1, l}\}$ and $\{E^c_{0, l}\}$. Let $S^0_k= A_0\cap S_k$  and $S^1_k=  A_1\cap S_k$ be the set of nulls and non-nulls remaining at iteration k. The relationship below always holds:
\[
\hat{\tau}^2_{S_{k}} \geq \frac{[|A_0|\tau^2_* - \sum_{i\in A_0\setminus S^0_k} x^2_i+ \sum_{i\in S^1_k} (x^2_i - \hat{\sigma}^2_i)]_+}{|S^0_{k}| + |S^1_{k}|}
\]
In other words, 
\begin{align*}
\{E^c_{k,l}, E_{k-1,l},   R_{k}\leq l\} \subseteq \{ \frac{[|A_0|\tau^2_* -  \sum_{i\in A_0\setminus S^0_k} x^2_i+  \sum_{i\in S^1_k} (x^2_i - \hat{\sigma}^2_i)]_+}{|S^0_{k}| + |S^1_{k}| } < \tau^2_l,  E_{k-1,l},   R_{k}\leq l\}
\end{align*}
If $\tau^2_l \leq 0$, the above event will never happen, hence,
\begin{align*}
\{E^c_{k,l}, E_{k-1,l},   R_{k}\leq l\} & \subseteq \{  |A_0|\tau^2_* -  \sum_{i\in A_0\setminus S^0_k} (x^2_i-\tau^2_l)+  \sum_{i\in S^1_k} (x^2_i - \hat{\sigma}^2_i-\tau^2_l)<\tau^2_l|A_0|,  \tau^2_l>0, E_{k-1,l},   R_{k}\leq l\}
\end{align*}
When $E_{k-1,l}$ happens and when $R_{k} \leq  l$, the removed nulls $A_0\setminus S^0_k$ must be in the set $B_l:=B_{1,l}\cap B_{2,l}$ for the following reasons. $A_0\setminus S^0_k\subseteq B_{2,l}$ by definition. Now, we let $p_* := \frac{|J^1_k|\alpha_1}{N}$ be the cut-off for the rejected p-values for the set $J^1_k$. There are two possibilities : $p_* \leq \alpha_2$ or $p_* > \alpha_2$. We discuss them separately.
\begin{enumerate}
\item If $p_* \leq \alpha_2$, the rejected set from ITEB will be the set $J^1_k$ and $J^1_k$ contains at most $l$ non-null hypothesis. Suppose $J^1_k$ contains exactly $s_0$ null hypotheses. In order for any null hypothesis $i$ to be rejected, it must satisfy $p_i \leq \frac{(s_0+l)\alpha_1}{N}$, and we know there are $s_0$ of them. As a result, we know $s_0= |\{i\in A_0: p_i \leq \frac{(s_0+l)\alpha_1}{N}\}|$. When $E_{k-1, l}$ happens, we have $p_i \geq p_{i,l}$, thus $s_0 \leq |\{i\in A_0: p_{i,l} \leq \frac{(s_0+l)\alpha_1}{N}\}|$ or $p_{(s_0), l} \leq \frac{(s_0+l)\alpha_1}{N}$. As a result, we have $s_0 \leq \arg\max\{s: p_{(s), l} \leq \frac{s+l}{N}\alpha_1\}$ and $A_0\setminus S^0_k \subseteq B_{1,l}$.  Hence, $A_0\setminus S^0_k \subseteq B_l$.
\item If $p^* > \alpha_2$,  the rejected set from ITEB will be the set $J^2_k$. In this case, we can show that $B_{2,l}\subseteq B_{1,l}$. Because $p_{i,l}\leq p_i$, everything in $J_k^1$ will again be rejected if we replace $p_i$ with $p_{i,l}$, in other words, $|B_{1,l}|\geq |J^1_k|$. As a result, $p_{**}$, the new cut-off p-value for $B_{1,l}$, will be larger than $\alpha_2$: $p_{**} \geq \frac{|J^1_k|\alpha_1}{N} > \alpha_2$, which in turns lead to the fact that $B_{2,l}\subseteq B_{1,l}$ and $B_{l} = B_{2,l}$. Hence,  $A_0\setminus S^0_k \subseteq B_l$.
\end{enumerate}
Let $\tilde{A}_{1, l} := \{i\in A_1: p_{i,l} \geq \alpha_2\}$. When $E_{k-1, l}$ holds,  $\tilde{A}_{1, l}\subseteq S^1_k$, and for any $i\in B_l$ or $i\in A_1\setminus \tilde{A}_{1, l}$, we have $x^2_i \geq \tilde{F}_i(\alpha_2)(\tau^2_l + \hat{\sigma}^2_i)$, thus  $ (x^2_i-\hat{\sigma}^2_i) > \tau^2_l$, and
\begin{align*}
\{E^c_{k,l}, E_{k-1,l},   R_{k}\leq l\} & \subseteq \{  |A_0|\tau^2_* -  \sum_{i\in B_l} (x^2_i-\tau^2_l)+  \sum_{i\in \tilde{A}_{1,l}} (x^2_i - \hat{\sigma}^2_i-\tau^2_l)<\tau^2_l|A_0|,  \tau^2_l>0\}
\end{align*}
We can construct $M$ based on the equation above.  We define the following six events:
\begin{align*}
& M_1= \{\forall l = 0, 1, \ldots, |A_1|: |B_{1, l}|\leq \frac{(\log^2 N\vee l)}{N}|A_0|\},  M_2 = \{\forall l = 0, 1, \ldots, |A_1|: |B_{2, l}| \leq 2\alpha_2|A_0|\}\\
&M_3= \{\forall l = 0,1,\ldots, |A_1|: (\max_{A_{\alpha}\subseteq A_0:|A_{\alpha}| \leq t_l|A_0|} \sum_{i\in A_{\alpha}}x^2_i) \leq 2.5(\tau^2+C)t_l|A_0|\log \frac{1}{t_l}\}\\
&M_{4}= \{\forall l = 0, 1, \ldots, |A_1|: \sum_{i\in \tilde{A}_{1,l}} (x^2_i-\hat{\sigma}^2_i-\tau^2_l)\geq -(\tau^2+1)(1-\gamma)\sqrt{|A_1|\log N}\}\\
&M_5= \{\sum_{i\in A_1}(x^2_i-\hat{\sigma}^2_i-\tau^2)\geq -(\tau^2+1)(1-\gamma)\sqrt{|A_1|\log N}\},\;\; M_0 = \{|\tau^2 - \tau^2_*|\leq (1-\sqrt{\gamma})\Delta_1 \}
\end{align*}
Let $M = \cap^5_{j=0}M_j$. Lemma \ref{lem:tailBounds} states that $M$ happens with probability approaching 1, whose proof is deferred to section \ref{app:lem}.
\begin{lemma}
\label{lem:tailBounds}
Under Assumption \ref{ass:ass1} and Assumption \ref{ass:ass2}, with $\alpha_1 <\frac{1}{2e}$ being a positive constant and $\alpha_2\rightarrow 0$ at a slow rate ($\alpha_2\frac{N}{\log^2 N}$ is bounded away from 0), $M_0$, $M_1$, $M_2$, $M_3$, $M_4$ and $M_5$ happen with probability approaching 1.
\end{lemma}
When $M_1$ and $M_2$ hold, we have $|B_l| = |B_{l,1}\cap B_{2,l}| \leq t_l |A_0|$, and if $M_3$ also holds, we have 
\[ 
\sum_{i\in B_l} x^2_i \leq \max_{A_{\alpha}\in A_0, |A_{\alpha}| \leq t_l|A_0|}\sum_{i\in A_{\alpha}} x^2_i \leq 2.5(\tau^2+C)|A_0|t_l  \log(\frac{1}{t_l}) \leq \frac{5}{6}|A_0| \Delta_{2,l}
\]
When $M_5$ holds, we have $\sum_{i\in \tilde{A}_{1,l}} (x^2_i -\hat{\sigma}^2_i -\tau^2_l ) \geq -\sqrt{\gamma}|A_0|\Delta_1$. Therefore, the following is true,
\begin{align*}
\{E^c_{k}, E_{k-1},   R_{k}\leq l, M\} &\subseteq  \{ |A_0|(\tau^2-(1-\sqrt{\gamma})\Delta_1) - \frac{5}{6}|A_0| \Delta_{2,l}-\sqrt{\gamma}|A_0|\Delta_1< |A_0|\tau^2_l, \tau^2_l>0\}\\
& =  \{(\tau^2-(1-\sqrt{\gamma})\Delta_1) - \frac{5}{6}\Delta_{2,l}-\sqrt{\gamma}\Delta_1<(\tau^2-\Delta_1-\Delta_{2,l})\}\\
& =  \{\frac{5}{6}\Delta_{2,l}>\Delta_{2,l}\} = \emptyset
\end{align*}
In the step 0, we use all points to estimate $\hat{\tau}^2_{S_0} = \frac{|A_0|\tau^2_*+\sum_{i\in A_1}(x^2_i - \hat{\sigma}^2_i)}{N}$. When $M_0$ holds, we have $|A_0|\tau^2_*\geq |A_0|(\tau^2-(1-\sqrt{\gamma})\Delta_1)$; when $M_5$ holds, we have $\sum_{i\in A_1} (x^2_i-\hat{\sigma}^2_i - \tau^2 ) \geq -|A_0|\sqrt{\gamma}\Delta_1$, thus we have $\{E^c_{0,l}, M\}= \{\hat{\tau}^2_{S_0} < [\tau^2 -\Delta_1-\Delta_{2,0}]_+, M\} = \emptyset$. 
\end{proof}
\subsection*{Proof of Theorem ~\ref{thm:upper}}
\begin{proof}
As $\hat{\tau}^2_{S_k}$ is non-decreasing, and in order for a point to be removed at any iteration $k$, it must be greater than $\hat{\tau}^2_{S_K}$: $x^2_i \geq \tilde{F}^{-1}_i(\alpha_2)(\hat{\tau}^2_{S_k}+\hat{\sigma}^2_i)\Rightarrow x^2_i-\hat{\sigma}^2_i > \hat{\tau}^2_{S_k} \geq \hat{\tau}^2_{S_K}$. Let $S^0_K = S_K\cap A_0$,  we have
\[
\sum_{i\in S^0_K}(x^2_i - \hat{\sigma}^2_i) \leq |A_0|\tau^2_* - (|A_0| - |S^0_K|)\hat{\tau}^2_{S_K}
\]
For  a point $i\in A_1$, in order for it to not be removed, it need to  satisfy the following criterion:
\[
x^2_i \leq \tilde{F}^{-1}_i(\alpha_2)(\hat{\sigma}^2_i+\hat{\tau}^2_{S_k}) \text{  or }  x^2_i \leq \tilde{F}^{-1}_i(\frac{l_i\alpha_1}{N})(\hat{\sigma}^2_i+\hat{\tau}^2_{S_k}) 
\]
where $l_i$ is the order of the p-value of $x^2_i$. As a result, at the last iteration $K$, we have
\[
\hat{\tau}^2_{S_K} \leq [\frac{|A_0|\tau^2_*-(|A_0| - |S_{K_0}|)\hat{\tau}^2_{S_K}}{|S_{K}|}+\frac{\sum_{i\in A_1}\lambda((\frac{l_i}{N}\alpha_1)\wedge\alpha_2)(\hat{\sigma}^2_i+\hat{\tau}^2_{S_K} )}{|S_{K}|}]_+
\]
If $\hat{\tau}^2_{S_K} = 0$, we have proved our statement; otherwise, the term insider the positive operator is positive, hence,  for $\hat{\tau}^2_{S_K} >0$, we have
\begin{align}
\label{eq:thm1_final}
(|A_0| - \sum_{i\in A_1}\lambda((\frac{l_i}{N}\alpha_1)\wedge \alpha_2))\hat{\tau}^2_{S_K}\leq |A_0| \tau^2_*+\sum_{i\in A_1}\lambda((\frac{l_i}{N}\alpha_1)\wedge \alpha_2) \hat{\sigma}^2_i 
\end{align}
We know that $\hat{\sigma}^2_i$ is $\frac{\chi^2_m}{m}\sigma^2_i$-distributed with mean at most $C$ and the variance at most $\frac{C}{m}$.  Apply the the Chebyshev's inequality  to the quantity $\sum_{i\in i\in A}a_i\hat{\sigma}^2_i$ for set $A$ and coefficient sequence $\{a_i\}$  we have
\begin{equation}
\label{eq:bound3}
P(\sum_{i\in A_1}a_i\hat{\sigma}^2_i \leq C(\sum_{i\in A_1} a_i+\sqrt{\frac{\sum_{i\in A_1} a^2_i\log N}{m}})) \rightarrow 1
\end{equation}
Based on Lemma \ref{lem:tailDecays}, there is a constant $c$ large enough such that for all $l = 1, 2, \ldots, |A_1|$, we have $\lambda(\frac{l\alpha_1}{N}) \leq c  (\frac{N}{\alpha_1 l})^{\frac{2}{5}}$, and 
\begin{align*}
&\sum^{ |A_1|}_{l = 1}\lambda(\frac{l\alpha_1}{N}) \leq c\sum^{|A_1|}_{l=1} (\frac{N}{l\alpha_1})^{\frac{2}{5}} \leq c (\frac{N}{\alpha_1})^{\frac{2}{5}}\int^{N\gamma}_0 l^{-\frac{2}{5}} d l = \frac{5c}{3\alpha_1^{\frac{2}{5}}}N \gamma^{\frac{3}{5}}\\
&\sum^{ |A_1|}_{l = 1}\lambda^2(\frac{l\alpha_1}{N}) \leq c^2\sum^{|A_1|}_{l=1} (\frac{N}{l\alpha_1})^{\frac{4}{5}} \leq c^2 (\frac{N}{\alpha_1})^{\frac{4}{5}}\int^{N\gamma}_0 l^{-\frac{4}{5}} d l = \frac{5c^2}{\alpha_1^{\frac{4}{5}}}N \gamma^{\frac{1}{5}}
\end{align*}
Combine the above inequality with equation (\ref{eq:bound3}), we have
\begin{align*}
&P\left(\sum_{i\in A_1}\lambda((\frac{l_i}{N}\alpha_1)\vee \alpha_2) \hat{\sigma}^2_i  < C\left( \frac{5c}{3\alpha_1^{\frac{2}{5}}}N \gamma^{\frac{3}{5}}+\gamma N\lambda(\alpha_2)+\sqrt{\frac{(5c^2\alpha_1^{-\frac{4}{5}}N \gamma^{\frac{1}{5}}
+N\gamma \lambda^2(\alpha_2))}{m}\log N }\right)\right)\rightarrow 1
\end{align*}
Let $c_1:= C\left( \frac{5c}{3\alpha_1^{\frac{2}{5}}}N \gamma^{\frac{3}{5}}+\gamma N\lambda(\alpha_2)+\sqrt{\frac{(5c^2\alpha_1^{-\frac{4}{5}}N \gamma^{\frac{1}{5}}
+N\gamma \lambda^2(\alpha_2))}{m}\log N }\right)$. Recall that $M_0 = \{|\tau^2-\tau^2_*| \leq (1-\sqrt{\gamma})\Delta_1\}$ happens with probability approaching 1 from Lemma \ref{lem:tailBounds}. For any $\delta >0$,
we have 
\begin{align*}
&\lim_{N\rightarrow \infty}P(\hat{\tau}^2 \leq \tau^2+\delta(\tau^2+C), M_0) \\
\leq& \lim_{N\rightarrow \infty}P(\frac{(1-\gamma)(\tau^2+\Delta_1)+ c_1}{1-\gamma\lambda(\alpha_2)- \frac{5a}{3}\gamma^{\frac{3}{5}}\alpha_1^{-\frac{2}{5}}} < \tau^2+\delta(\tau^2+C)) - \lim_{N\rightarrow \infty}P(M^c_0)-\lim_{N\rightarrow \infty}P(\sum_{i\in A_1}\lambda((\frac{l_i}{N}\alpha_1)\vee \alpha_2) \hat{\sigma}^2_i > c_1)\\
= & \lim_{N\rightarrow \infty}P(\frac{(1-\gamma)(\tau^2+\Delta_1)+ c_1}{1-\gamma\lambda(\alpha_2)- \frac{5a}{3}\gamma^{\frac{3}{5}}\alpha_1^{-\frac{2}{5}}} <\tau^2+\delta(\tau^2+C))\\
= & \lim_{N\rightarrow \infty}P(\frac{(\gamma\lambda(\alpha_2)+ \frac{5a}{3}\gamma^{\frac{3}{5}}\alpha_1^{-\frac{2}{5}}-\gamma)\tau^2+(1-\gamma)\Delta_1+ c_1}{1-\gamma\lambda(\alpha_2)- \frac{5a}{3}\gamma^{\frac{3}{5}}\alpha_1^{-\frac{2}{5}}} <\delta(\tau^2+C)) = 1
\end{align*}
\end{proof}

\subsection*{Proof of  Theorem \ref{thm:FDR}}
\begin{proof}
Let $R_0$  and $R_1$ be the number of rejected nulls and non-nulls using level $\alpha_1$(note that $R_1$ is the $R_K^1$ in Theorem \ref{thm:lower}).  Define $V_i = \mathbbm{1}_{\{H_i\;rejected\}}$ for each $i\in A_0$,   $p_i$ be the p values calculated using $\hat{\tau}^2$. We can express the FDR as
\begin{align*}
FDR &=\sum^{|A_1|}_{l=0}\sum^{|A_0|}_{l_0 = 1}E[\mathbbm{1}_{R_1 = l} \mathbbm{1}_{R_0 = l_0} \frac{\sum_{i\in A_0} V_i}{l+l_0}] =\sum^{|A_1|}_{l=0}\sum^{|A_0|}_{l_0 = 1}E[\mathbbm{1}_{R_1 = l} \mathbbm{1}_{R_0 = l_0} \frac{\sum_{i\in A_0} \mathbbm{1}_{p_i \leq (\frac{l+l_0}{N}\alpha_1)\wedge \alpha_2}}{l+l_0}]  
\end{align*}
We can further decompose the expression for FDR into two parts
{\small
\[
FDR = \underbrace{\sum^{|A_1|}_{l=0}\sum^{|A_0|}_{l_0 = 1}E[\mathbbm{1}_{R_1 = l} \mathbbm{1}_{R_0 = l_0} \mathbbm{1}_{\hat{\tau}^2 \geq \tau^2_l} \frac{\sum_{i\in A_0} \mathbbm{1}_{p_i \leq (\frac{l+l_0}{N}\alpha_1)\wedge \alpha_2}}{l+l_0}]  
}_{I_{1}}+\underbrace{\sum^{|A_1|}_{l=0}\sum^{|A_0|}_{l_0 = 1}E[\mathbbm{1}_{R_1 = l} \mathbbm{1}_{R_0 = l_0} \mathbbm{1}_{\hat{\tau}^2 < \tau^2_l} \frac{\sum_{i\in A_0} \mathbbm{1}_{p_i \leq (\frac{l+l_0}{N}\alpha_1)\wedge \alpha_2}}{l+l_0}] }_{I_{2}}
\]
}
By Theorem \ref{thm:lower}, we know $I_{2} \leq P(\cup^{|A_1|}_{l=0}\{R_1 = l, \hat{\tau}^2 < \tau^2_l\})\rightarrow 0$, and we need only to bound $I_{1}$.  Let $\mathcal{F}_i = \{x^2_1, \ldots, x^2_{i-1}, x^2_{i+1}, \ldots, x^2_N,\hat{\sigma}^2_1, \ldots, \hat{\sigma}^2_{i-1}, \hat{\sigma}^2_{i+1}, \ldots, \hat{\sigma}^2_N \}$.  Notice that
\begin{itemize}
\item Let us take $x^2_i$ and $\hat{\sigma}^2_i$ and set their value to $\infty$ and $0$, and denote new number of rejections for the null and non-null by $\tilde{R}_0$ and $\tilde{R}_1$. If $p_i$ is rejected, we know $\hat{\tau}^2$ is not calculated using $x^2_i$ or $\hat{\sigma}^2_i$. This new number of rejections is exactly $R_0$ and $R_1$ if we have rejected hypothesis $i$:
\[
\{R_1 = l,  R_0 = l_0, p_{i} \leq \frac{\alpha_1(l+l_0)}{N}\wedge\alpha_2,  \hat{\tau}^2\geq \tau^2_l\} =\{\tilde{R}_1 = l,\tilde{R}_0= l_0,  p_{i} \leq \frac{\alpha_1(l+l_0)}{N}\wedge\alpha_2,   \hat{\tau}^2\geq \tau^2_l\}  
\]
\end{itemize}
We take the expectation conditional on $\mathcal{F}_i$:
\begin{align}
\label{eq:FDRI1}
I_{1}&=\sum_{i\in A_0}\sum^{|A_1|}_{l = 0} \sum^{|A_0|}_{l_0 = 1} E[\mathbbm{1}_{\tilde{R}_1 = l}   \mathbbm{1}_{\tilde{R}_0= l_0}  E[\frac{\mathbbm{1}_{p_{i} \leq \frac{\alpha_1(l+l_0)}{N}\wedge \alpha_2} }{(l+l_0)} \mathbbm{1}_{\hat{\tau}^2 \geq \tau^2_l}|\mathcal{F}_i]] \nonumber \\
&\leq  \sum_{i\in A_0}\sum^{|A_1|}_{l = 0} \sum^{|A_0|}_{l_0 = 1} E[\mathbbm{1}_{\tilde{R}_1 = l}   \mathbbm{1}_{\tilde{R}_0= l_0} E[\frac{\mathbbm{1}_{p_{i,l} \leq \frac{\alpha_1(l+l_0)}{N}\wedge \alpha_2} }{(l+l_0)}|\mathcal{F}_i]]\nonumber \\
 &=\sum_{i\in A_0} \sum^{|A_1|}_{l = 0} \sum^{|A_0|}_{l_0 = 1} E[\mathbbm{1}_{\tilde{R}_1 = l}   \mathbbm{1}_{\tilde{R}_0= l_0} E[\frac{\mathbbm{1}_{p_{i,l} \leq \frac{\alpha_1(l+l_0)}{N}\wedge \alpha_2} }{(l+l_0)}]]
\end{align}
By Lemma \ref{lem:tailDecays}, part (2), we know $\lim_{N\rightarrow \infty}\sup_{i\in A_0}\sup_{l\geq 0}\sup_{\alpha \geq\min\left(\alpha_2, \frac{(l\vee 1)\alpha_1}{N}\right)}  \frac{P(p_{i,l} \leq \alpha)}{\alpha} = 1$.  As a result, for any $\delta >0$, there exists a $N_0$ such that for all $N > N_0$, we have
\[
\sup_{l\geq 0}\sup_{l_0 \geq 1} P(p_{i,l} \leq (\frac{l+l_0}{N} \alpha_1)\wedge \alpha_2) \leq (1+\delta)(\frac{l+l_0}{N}\alpha_1)\wedge\alpha_2
\]
Rearrange the righthand side of equation (\ref{eq:FDRI1}), we have $I_{1} \leq (1+\delta)\alpha_1$ for any $\delta > 0$. Hence $\lim_{N\rightarrow\infty}I_{1,1} \leq \alpha_1$ and $\lim_{N\rightarrow\infty}FDR \leq \alpha_1$.
\end{proof}

\subsection*{Proof of Theorem \ref{thm:optimality}}

\begin{proof}
From Theorem \ref{thm:upper}, for any $\delta_1 > 0$, we know $M =\{\hat{\tau}^2 \leq \tau^2+\delta_1(\tau^2+C)\}$ happens with probability approaching 1, which leads to the following result:
\begin{align*}
P(\phi_{i, \alpha} = 0)  \leq &P(x^2_i \leq \widetilde{F}^{-1}_i(\alpha)(\hat{\tau}^2+\hat{\sigma}^2_i), M)+P(M^c) \\
\leq &\underbrace{P(z^2_i \leq \frac{\widetilde{F}^{-1}_i(\alpha)(\tau^2+\hat{\sigma}^2_i+\delta_1(\tau^2+C))}{\tau^2_i +\sigma^2_i})}_{I_{i,\alpha}} + P(M^c)
\end{align*}
We now prove that $I_{i,\alpha}$ is no much larger than the oracle loss. We know that there exists a constant $f_{max} \geq \sup_t \frac{d P(z^2_i\leq t)}{dt }$, and for any $\delta_2$, there is a constant $w$ large enough such that  
\[
\sup_{\delta > 0}\sup_{i\in A_1} \frac{P(z^2_i \leq w(1+\delta))}{P(z^2_i \leq w)(1+\delta)}\leq 1+\delta_2
\]
For any $\alpha$, we either have $\frac{\tilde{F}^{-1}_i(\alpha)(\tau^2+C)}{\tau^2_i+\sigma^2_i}\leq \frac{w}{\sqrt{\delta_1}}$ or not.  If $\frac{\tilde{F}^{-1}_i(\alpha)(\tau^2+C)}{\tau^2_i+\sigma^2_i}\leq \frac{w}{\sqrt{\delta_1}}$, we have $I_{i,\alpha} \leq P(\phi^*_i = 0)+wf_{max}\sqrt{\delta_1}$. If$\frac{\tilde{F}^{-1}_i(\alpha)(\tau^2+C)}{\tau^2_i+\sigma^2_i}> \frac{w}{\sqrt{\delta_1}}$,  we have
\begin{align*}
I_{i,\alpha}\leq &\int^{\infty}_{\sqrt{\delta_1}C}  P(z^2_i \leq \frac{\widetilde{F}^{-1}_i(\alpha)(\tau^2+y)}{\tau^2_i +\sigma^2_i})(\frac{\tau^2+y+\delta_1(\tau^2+C)}{\tau^2+y})f_{\hat{\sigma}^2_i}(y) dy + P(\hat{\sigma}^2_i \leq \sqrt{\delta_1}C)+\delta_2\\
\leq &P(\phi^*_{i,\alpha} = 0)+\underbrace{\int^{\infty}_{0} P(\bar{x}^2_i \leq \widetilde{F}^{-1}_i(\alpha)(\tau^2+y))\frac{\delta_1(C+\tau^2)}{y+\tau^2}f_{\hat{\sigma}^2_i}(y) dy}_{I_1}+P(\hat{\sigma}^2_i \leq \delta_1 \sigma^2_i)+\delta_2
\end{align*}
In the integral $I_1$, because $P(x^2_i \leq \widetilde{F}^{-1}_i(\alpha)(\tau^2+\hat{\sigma}^2_i+y))$ is an increasing function in $y$ while $\frac{\delta_1(C+\tau^2)}{y+\tau^2}$ is a decreasing function in $y$, we have
\[
I_1 \leq  P(\phi^*_i = 0) \int^{\infty}_{0} \frac{\delta_1(C+\tau^2)}{y+\tau^2}f_{\hat{\sigma}^2_i}(y) dy 
\]
$\hat{\sigma}^2_i$ is $\frac{\chi^2_{m}}{m\sigma^2_i}$ distributed, the expectation of its inverse is $\frac{m}{(m-2)}$, Recall that $\min \sigma^2_i = 1$:
\[
\int^{\infty}_{0} \frac{\delta_1(C+\tau^2)}{y+\tau^2}f_{\hat{\sigma}^2_i}(y) dy \leq \int^{\infty}_{0} (\frac{\delta_1C}{y}+\delta_1)f_{\hat{\sigma}^2_i}(y) dy =\delta_1(1+\frac{Cm}{m-2})
\]
As a result, we have
\[
P(\phi_{i, \alpha} = 0) -P(\phi^*_i = 0) \leq \max_{i\in A_1}(\sqrt{\delta_1}wf_{max}, \delta_1+\delta_1\frac{Cm}{m-2}+P(\hat{\sigma}^2_i \leq \sqrt{\delta_1}C) +\delta_2)+P(M^c)
\]
The left-hand-side of  the above expression does not depend of $i$ or $\alpha$. For any $\delta > 0$, we can take $N$ large enough and $\delta_1$, $\delta_2$ small enough  such that 
\[
\delta_1+\delta_1\frac{Cm}{m-2}+\max_{i\in A_1} P(\hat{\sigma}^2_i \leq \sqrt{\delta_1}C) +\delta_2+P(M^c)< \delta, \;\;\sqrt{\delta_1}wf_{max}  +P(M^c)< \delta
\]
Hence, we have $ \lim_{N\rightarrow \infty}\sup_{i\in A_1}\sup_{\alpha\geq 0}(P(\phi_{i, \alpha}= 1) - P(\phi^*_{i, \alpha}= 1)) \geq 0$.
\end{proof}

\section{Proof of Lemmas \ref{lem:tailDecays},  \ref{lem:tailBounds} }
\label{app:lem}
\subsection*{Proof of Lemma \ref{lem:tailDecays}}
\begin{proof}
(1) Let $1-\tilde{T}_m(.)$ be the cumulative function of a t distribution with $m$ degree of freedom and $1-\tilde{\Phi}(.)$ be the cumulative function of a normal. Let $t_m(.)$ and $\phi(.)$ be there density function. We first show that for any fixed value $t\geq 0$, we have
\begin{equation}
\label{eq:tailbound}
2\tilde{\Phi}(\sqrt{t}) \leq \tilde{F}_i(t) \leq 2\tilde{T}_m(\sqrt{t})
\end{equation}
Let $a(\tau^2) :=P(\frac{x^2_i}{\tau^2+\hat{\sigma}^2_i} \geq t) = P(\frac{(\tau^2+\sigma^2_i)z^2}{\tau^2+\sigma^2_iu} \geq t)$,  where $z$ be a random variable with standard normal distribution and $u$ be a random variable distributed as $\frac{\chi^2_m}{m}$, $z$ and $u$ are independent. The function $a(\tau^2)$ has a non-positive first derivative with respect to $\tau^2$:
\begin{align*}
\frac{d a(\tau^2)}{d \tau^2} &= \frac{d}{d \tau^2} \int^{\infty}_{0}\int_{z^2\geq \frac{t(\tau^2+u\sigma^2_i)}{\tau^2+\sigma^2_i}} \phi(z) dz f_u(u) d u\\
 &=\int^\infty_{u=0} f_u(u) \phi(\sqrt{\frac{t(\tau^2+u\sigma^2_i)}{\tau^2+\sigma^2_i}})\sqrt{\frac{\tau^2+\sigma^2_i}{t(\tau^2+u\sigma^2_i) }}\frac{\sigma^2_it(u-1)}{(\tau^2+\sigma^2_i)^2} du\\
& \propto \int^\infty_{u=0} f_u(u) e^{-\frac{t(\tau^2+u\sigma^2_i)}{\tau^2+\sigma^2_i}}\sqrt{\frac{1}{\tau^2+u\sigma^2_i }}(u-1) du
\end{align*}
The expected value of $u$ is 1: $\int^\infty_{x=0} f_u(x)(u-1) = 0$ and $e^{-\frac{t(\tau^2+u\sigma^2_i)}{\tau^2+\sigma^2_i}}\sqrt{\frac{1}{\tau^2+u\sigma^2_i }}$ is a decreasing function of $u$, thus $a(\tau^2)$ has a non-positive first derivative with respect to $\tau^2$. For any fixed $t\geq 0$, we have
\[
a(\infty)\leq P(\frac{x^2_i}{\tau^2+\hat{\sigma}^2_i} \geq t)  \leq a(0)
\]
We use the fact that $a(\infty) = 2\tilde{\Phi}(\sqrt{t})$ and $a(0) = 2\tilde{T}_m(\sqrt{t})$ to get equation (\ref{eq:tailbound}). It is also easy to check that for any fixed non-negative $t$, $\tilde{T}_m(t)$ is non-increasing in $m$ because when $m_1 < m_2$, the density ratio between the t-distribution with degree of freedom $m_1$ and that with degree of freedom $m_2$  is non-decreasing in the positive part and non-increasing in the negative part. As a result, $\tilde{T}_m(t)$ is non-increasing in $m$ for any fixed $t$ and $\tilde{F}_i(t) \leq 2\tilde{T}_5(\sqrt{t})$. Apply the Mill's ratio result for the t-distribution(\cite{soms1976asymptotic}):
\begin{equation}
\label{eq:Millt}
\tilde{T}_m(t) < \frac{t_m(t)}{t}(1+\frac{t^2}{m})
\end{equation}
we have
\[
\tilde{F}(t)\leq \frac{2\Gamma(\frac{m+1}{2})}{\sqrt{\pi mt}\Gamma(\frac{m}{2})}(1+\frac{t}{m})^{-\frac{m-1}{2}} < \sqrt{\frac{2}{\pi}}(\frac{t}{m})^{-\frac{m}{2}} \rightarrow \lambda(\frac{l\alpha_1}{N}) \lesssim (\frac{N}{l\alpha_1})^{\frac{2}{5}}
\]
As a direct result, we have $\frac{\Delta_1}{\tau^2+1}\lambda(\frac{\alpha_1}{N}) \rightarrow 0$.  Because $\Delta_{2,l }\leq \alpha_2 \log(\frac{1}{\alpha_2})$, we have $\lambda(\alpha_2) \alpha_2 \log\frac{1}{\alpha_2}\rightarrow 0$, hence $\sup_{l\geq 0}\Delta_{2,l}\lambda(\lambda_2)\rightarrow 0$. The result $\sup_{l\geq 1}\frac{\Delta_{2,l}}{\tau^2+1}\lambda(\frac{l\alpha_1}{N}) \rightarrow 0$ also holds because
\begin{itemize}
\item If $\frac{l}{N}$ is a positive constant, $\frac{\Delta_{2,l}}{\tau^2+1}\rightarrow 0$ because $\alpha_2\rightarrow 0$.
\item If $\frac{l}{N}\rightarrow 0$ and $\frac{l}{N} \gtrsim \frac{\log^2 N}{N}$, $\frac{\Delta_{2,l}}{\tau^2+1} \lesssim \frac{l}{N}\log \frac{N}{l}$, $\lambda(\frac{l\alpha_1}{N})\frac{\Delta_{2,l}}{\tau^2+1}\rightarrow 0$.
\item If $\frac{l}{N}\lesssim \frac{\log^2 N}{N}$, $\frac{\Delta_{2,l}}{\tau^2+1} \lesssim \frac{\log^2 N}{N}\log\frac{N}{\log^2 N}$ and $N^\frac{2}{5}\frac{\Delta_{2,l}}{\tau^2+1}\rightarrow 0$, hence we still have $\lambda(\frac{l\alpha_1}{N})\frac{\Delta_{2,l}}{\tau^2+1}\rightarrow 0$.
\end{itemize}
(2)Based on  part (1) and the fact that $\Delta_{2,0}\leq \Delta_{2,1}$, let $\alpha_3 = \min\left(\alpha_2, \frac{(l\vee1)\alpha_1}{N}\right)$ , we have:
\begin{equation}
\label{eq:diffbound}
\sup_{l\geq 0}\sup_{\alpha\geq  \alpha_3 }\frac{\Delta_1 +\Delta_{2,l}}{\tau^2+1}\lambda(\alpha) \rightarrow 0 \end{equation}
Because $\tau^2_l \leq \tau^2$, we always have $\frac{P(p_{i,l} \leq \alpha)}{\alpha} \geq 1$, and we need only to check that, for any $\delta > 0$, $\frac{P(p_{i,l} \leq \alpha)}{\alpha} \leq 1+\delta$ holds uniformly for large $N$.
We break the expression in the statement  into two parts:
\begin{align*}
&\sup_{i,l}\sup_{\alpha\geq \alpha_3}P(p_{i,l} \leq \alpha))/\alpha = \sup_{i,l, \alpha}\int^{\infty}_{y = 0} P(x^2_i\geq (y+\tau^2_l)\tilde{F}^{-1}_i(\alpha)) f_{\hat{\sigma}^2_i}(y)dy/\alpha =\sup_{i,l,\alpha}(I_{1,i,l,\alpha} + I_{2,i,l,\alpha})
\end{align*}
where $I_{1, i, l,\alpha} =\int_{\frac{y+\tau^2}{\tau^2+C}\tilde{F}_i^{-1}(\alpha) > 
\frac{1}{\delta}} P(x^2_i\geq (y+\tau^2_l)\tilde{F}^{-1}_i(\alpha))f_{\hat{\sigma}^2_i}(y) dy/\alpha$ and $I_{2,i,l,\alpha} = \int_{\frac{y+\tau^2}{\tau^2+C}\tilde{F}_i^{-1}(\alpha) \leq
\frac{1}{\delta}} P(x^2_i\geq (y+\tau^2_l)\tilde{F}^{-1}_i(\alpha))f_{\hat{\sigma}^2_i}(y) dy/\alpha$, with $\delta$ being any positive constant. For $I_{1, i, l,\alpha}$, we use the following Mill's result for the normal(\cite{gordon1941values})
\begin{equation}
 \label{eq:Millnorm}
 \frac{t}{t^2+1}\phi(t) < \tilde{\Phi}(t)<\frac{\phi(t)}{t}
\end{equation}
to upper bound $ P(x^2_i\geq (y+\tau^2_l)\tilde{F}^{-1}_i(\alpha))$ and lower bound $ P(x^2_i\geq (y+\tau^2)\tilde{F}^{-1}_i(\alpha))$ in terms of the density:
\begin{align*}
&I_{1, i, l, \alpha} \leq   \int_{\frac{y+\tau^2}{\tau^2+C}\tilde{F}_i^{-1}(\alpha)> 
\frac{1}{\delta}} P(x^2_i\geq (y+\tau^2)\tilde{F}^{-1}_i(\alpha))\frac{\tau^2+\sigma^2_i+(y+\tau^2)\tilde{F}^{-1}_i(\alpha)}{(y+\tau^2_l)\tilde{F}_i^{-1}(\alpha)}e^{\frac{(\Delta_1+\Delta_{2,l})\tilde{F}^{-1}_i(\alpha)}{2(\tau^2+\sigma^2_i)}}f_{\hat{\sigma}^2_i}(y) dy/\alpha\\
& \leq \int_{\frac{y+\tau^2}{\tau^2+C}\tilde{F}_i^{-1}(\alpha)> 
\frac{1}{\delta}} P(x^2_i\geq (y+\tau^2)\tilde{F}^{-1}_i(\alpha))(1+\frac{\tau^2+\sigma^2_i+(\Delta_1+\Delta_{2,l})\tilde{F}^{-1}_i(\alpha)}{(y+\tau^2_l)\tilde{F}_i^{-1}(\alpha)})e^{\frac{(\Delta_1+\Delta_{2,l})\tilde{F}^{-1}_i(\alpha)}{2(\tau^2+\sigma^2_i)}}f_{\hat{\sigma}^2_i}(y) dy/\alpha\\
\end{align*}
For $I_{2,i,l, \alpha}$:
\begin{align*}
&I_{2,i,l, \alpha} - \int_{\frac{y+\tau^2}{\tau^2+C}\tilde{F}_i^{-1}(\alpha) \leq \frac{1}{\delta}} P(x^2_i\geq (y+\tau^2)\tilde{F}^{-1}_i(\alpha))f_{\hat{\sigma}^2_i}(y) dy/\alpha\\
=& \int_{\frac{y+\tau^2}{\tau^2+C}\tilde{F}_i^{-1}(\alpha) \leq \frac{1}{\delta}} \left(P(x^2_i\geq (y+\tau^2_l)\tilde{F}^{-1}_i(\alpha)) - P(x^2_i\geq (y+\tau^2)\tilde{F}^{-1}_i(\alpha))\right)f_{\hat{\sigma}^2_i}(y) dy/\alpha 
\end{align*}
Recall that  $P(x^2_i\geq (y+\tau^2)\tilde{F}^{-1}_i(\alpha)) = 2\tilde{\Phi}(\sqrt{\frac{(y+\tau^2)\tilde{F}^{-1}_i(\alpha)}{\tau^2+\sigma^2_i}})$ and $P(x^2_i\geq (y+\tau^2_l)\tilde{F}^{-1}_i(\alpha)) = 2\tilde{\Phi}(\sqrt{\frac{(y+\tau^2_l)\tilde{F}^{-1}_i(\alpha)}{\tau^2+\sigma^2_i}})$, we can bound the difference by the product of the difference in the interval length and the upper bound of the normal density
\[
2\tilde{\Phi}(\sqrt{\frac{(y+\tau^2_l)\tilde{F}^{-1}_i(\alpha)}{\tau^2+\sigma^2_i}}) - 2\tilde{\Phi}(\sqrt{\frac{(y+\tau^2)\tilde{F}^{-1}_i(\alpha)}{\tau^2+\sigma^2_i}})\leq 2\sqrt{\frac{1}{2\pi}}(\sqrt{\frac{(y+\tau^2)\tilde{F}^{-1}_i(\alpha)}{\tau^2+\sigma^2_i}} - \sqrt{\frac{(y+\tau^2_l)\tilde{F}^{-1}_i(\alpha)}{\tau^2+\sigma^2_i}})
\]
We know that for any positive value $x, y$, we have $\sqrt{x+y} \leq \sqrt{x}+\sqrt{y}$, as a result, we have
\[
\frac{2\tilde{\Phi}(\sqrt{\frac{(y+\tau^2_l)\tilde{F}^{-1}_i(\alpha)}{\tau^2+\sigma^2_i}}) - 2\tilde{\Phi}(\sqrt{\frac{(y+\tau^2)\tilde{F}^{-1}_i(\alpha)}{\tau^2+\sigma^2_i}})}{2\tilde{\Phi}(\sqrt{\frac{(y+\tau^2)\tilde{F}^{-1}_i(\alpha)}{\tau^2+\sigma^2_i}})} \leq \frac{\sqrt{\frac{(\tau^2-\tau^2_l)\tilde{F}^{-1}_i(\alpha)}{2\pi(\tau^2+\sigma^2_i)}}}{\tilde{\Phi}(\sqrt{\frac{(y+\tau^2)\tilde{F}^{-1}_i(\alpha)}{\tau^2+\sigma^2_i}})}
\]
In other words, we have
\[
I_{2,i,l, \alpha} \leq  \int_{\frac{y+\tau^2}{\tau^2+C}\tilde{F}_i^{-1}(\alpha) \leq \frac{1}{\delta}} P(x^2_i\geq (y+\tau^2)\tilde{F}^{-1}_i(\alpha))(1+\frac{\sqrt{\frac{(\tau^2-\tau^2_l)\tilde{F}^{-1}_i(\alpha)}{2\pi(\tau^2+\sigma^2_i)}}}{\tilde{\Phi}(\sqrt{\frac{C}{\delta}})})f_{\hat{\sigma}^2_i}(y) dy/\alpha
\]
Combine them together and apply equation (\ref{eq:diffbound}), we have
\begin{align*}
&\sup_{i,l, \alpha}(I_{1,i,l, \alpha}+I_{2,i,l, \alpha} - 1)\leq   \sup_{i,l,\alpha}(\int_{\frac{y+\tau^2}{C+\tau^2}\tilde{F}^{-1}(\alpha) \leq \frac{1}{\delta}} P(x^2_i\geq (y+\tau^2)\tilde{F}^{-1}_i(\alpha)) \frac{\sqrt{\frac{(\Delta_1+\Delta_{2,l})\tilde{F}^{-1}(\alpha)}{2\pi(\tau^2+C)}}}{\tilde{\Phi}(\sqrt{\frac{C}{\delta}})}f_{\hat{\sigma}^2_i}(y) dy/\alpha\\ 
&+\int_{\frac{y+\tau^2}{C+\tau^2}\tilde{F}_i^{-1}(\alpha) > 
\frac{1}{\delta}} P(x^2_i\geq (y+\tau^2)\tilde{F}^{-1}_i(\alpha))\frac{\tau^2+\sigma^2_i+(\Delta_1+\Delta_{2,l})\tilde{F}^{-1}_i(\alpha)}{(y+\tau^2_l)\tilde{F}_i^{-1}(\alpha)}e^{\frac{(\Delta_1+\Delta_{2,l})\tilde{F}^{-1}_i(\alpha)}{2(\tau^2+\sigma^2_i)}}f_{\hat{\sigma}^2_i}(y) dy/\alpha)\\
&\rightarrow \sup_{i,l,\alpha} \int_{\frac{y+\tau^2}{C+\tau^2}\tilde{F}_i^{-1}(\alpha) > 
\frac{1}{\delta}} P(x^2_i\geq (y+\tau^2)\tilde{F}^{-1}_i(\alpha))\frac{\tau^2+\sigma^2_i}{(y+\tau^2_l)\tilde{F}_i^{-1}(\alpha)}f_{\hat{\sigma}^2_i}(y) dy/\alpha \\
&\leq \sup_{i,l,\alpha} \int_{\frac{y+\tau^2}{C+\tau^2}\tilde{F}_i^{-1}(\alpha) > 
\frac{1}{\delta}} P(x^2_i\geq (y+\tau^2)\tilde{F}^{-1}_i(\alpha))\frac{\tau^2+C}{\frac{(\tau^2+C)}{\delta}-(\Delta_1+\Delta_{2,l})\tilde{F}_i^{-1}(\alpha)}f_{\hat{\sigma}^2_i}(y) dy/\alpha \\
&\rightarrow \sup_{i,\alpha} \int_{\frac{y+\tau^2}{C+\tau^2}\tilde{F}_i^{-1}(\alpha) > 
\frac{1}{\delta}} \delta P(x^2_i\geq (y+\tau^2)\tilde{F}^{-1}_i(\alpha))f_{\hat{\sigma}^2_i}(y) dy/\alpha <\delta
\end{align*}
As it holds for any $\delta > 0$, we have
$\lim_{N\rightarrow \infty}\sup_{i\in A_0}\sup_{l\geq 0} \sup_{\alpha \geq \min\left((\alpha_2\vee\frac{\log^2 N}{N}), \frac{(l\vee1)\alpha_1}{N}\right) }  \frac{P(p_{i,l} \geq \alpha)}{\alpha} = 1$
 
(3) At ${k+1}^{th}$ iteration, for every point we removed, they need to satisfy that $x^2_i \geq \tilde{F}^{-1}_i(\alpha_1)(\hat{\tau}^2_{S_k}+\hat{\sigma}^2_i)$. From equation (\ref{eq:tailbound}),  we have(recall that $\alpha_1 < \frac{1}{2e}$):
\[
\tilde{F}^{-1}(\alpha_1) \geq (\tilde{\Phi}^{-1}(\frac{\alpha_1}{2}))^2   > 1
\]
as a result, $x^2_i - \hat{\sigma}^2_i \geq \hat{\tau}^2_{S_k}\Rightarrow$ the $\tau^2$ estimate is non-increasing.
\end{proof}

\subsection*{Proof of Lemma \ref{lem:tailBounds}}
\textbf{Proof of $M_0$ happening with probability approaching one:} We know that $\hat{\sigma}^2_i\sim \sigma^2_i\frac{\chi^2_m}{m}$ and $x^2_i\sim (\tau^2+\sigma^2_i)\chi^2_1$.  Because $\sigma^2_i\leq C$, we have $E[x^2_i - \hat{\sigma}^2_i] = \tau^2$  and  $Var[x^2_i - \hat{\sigma}^2_i] \leq \tau^2+(1+\frac{1}{m})C$. Result follows from the Chebyshev's inequality.

\noindent\textbf{Proof of $M_1$ happening with probability approaching one:} For the event $M_1$, consider the event $A_{k,l}$ := $\{|B_{1,l}| =  k\}$. Use Lemma \ref{lem:tailDecays} part (2),  and take $\delta < \frac{1}{2e\alpha_1}-1$, for large enough $N$:
\begin{equation}
\label{eq:count1}
\sup_{i\in A_0}\sup_{l\geq 0, k\geq 1} \frac{P(p_{i,l} \leq \frac{(l+k)\alpha_1}{N})}{\frac{(l+k)\alpha_1}{N}} < 1+\delta
\end{equation}
Event $A_{k,l}$ is contained in the event that there are  $k$  null p-values at most $\frac{(l+k)\alpha_1}{N}$, hence, $P(A_{k,l}) \leq \left(\begin{array}{c} |A_0|\\ k\end{array}\right)(\frac{l+k}{N}(1+\delta)\alpha_1)^k$.
Let $k_l := \lceil |A_0|\max(\frac{l}{N}, \frac{\log^2 N}{N})\rceil$, for $l = 0, 1, \ldots, |A_1|$, we have
\begin{align*}
P(M_1^c) \leq \sum^{|A_1|}_{l=0}\sum_{k \geq k_l} \left(\begin{array}{c} |A_0|\\ k\end{array}\right) (\frac{l+k}{N}(1+\delta)\alpha_1)^k
\end{align*}
Let $a_{k,l} = \left(\begin{array}{c} |A_0|\\ k\end{array}\right)(\frac{k+l}{N}(1+\delta)\alpha_1)^k$, $u_{k,l} = \frac{k+l}{k}$. It is easy to check that $u_{k,l}$ is decreasing in $k$ and $x\mapsto xe^{1/x}$ is increasing on $[1, \infty]$.  For $k \geq k_l\geq (1-\gamma)l$, we have that $u_{k,l} \leq \frac{2}{1-\gamma}$, and $(1+\frac{1}{k+l})^k \leq e^{1/u_{k,l}}$, as $\log (1+\frac{1}{k+l})^k = k \log (1+\frac{1}{k+l}) \leq \frac{k}{k+l}$. Hence, for large $N$ and any $l$ considered, $a_{k,l}$ is non-increasing in $k$ when $k \geq k_l$:
\begin{align*}
\sup_{l, k}\frac{a_{k+1,l}}{a_{k,l}} =&\sup_{l, k} \frac{(|A_0|-k)}{k+1}\frac{(k+l+1)(1+\delta)\alpha_1}{N}(1+\frac{1}{k+l})^{k}\\
\leq &\sup_{l, k} (1-\gamma)(1+\delta)\alpha_1 u_{k,l}e^{\frac{1}{u_{k,l}}} \\
 \leq&(1-\gamma)(1+\delta)\alpha_1 \frac{2}{1-\gamma}e^{\frac{1-\gamma}{2}} < 2\sqrt{e}(1+\delta)\alpha_1 < 1
\end{align*}
Using sterling's approximations to upper bound $a_{l,  k_l}$, the probability of $M^c_1$ can be bounded as
\begin{align*}
P(M_1^c)&\leq\sum^{|A_1|}_{l=0}|A_0| a_{l, \lceil k_l\rceil }\leq \sum^{|A_1|}_{l=0} |A_0|\frac{|A_0|^{k_l+|A_0|-k_l}}{k^{k_l}_l(|A_0| -k_l)^{|A_0| - k_l}}(\frac{k_l+l}{N}(1+\delta)\alpha_1)^{k_l} \\
& =\sum^{|A_1|}_{l=0} |A_0|\frac{|A_0|^{|A_0|-k_l}}{(|A_0| -k_l)^{|A_0| - k_l}}(\frac{|A_0|}{N})^{k_l}(\frac{k_l+l}{k_l}(1+\delta)\alpha_1)^{k_l}\\
&= \sum^{|A_1|}_{l=0} |A_0|\exp(k_l\log(u_{k_l, l}(1-\gamma)(1+\delta)\alpha_1)+(|A_0|-k_l)\log \frac{|A_0|}{|A_0| - k_l})\\
& \leq \sum^{|A_1|}_{l=0} |A_0|\exp\left(k_l\log(2(1+\delta)\alpha_1)+(|A_0|-k_l)\log \frac{|A_0|}{|A_0| - k_l}\right)
\end{align*}
The quantity inside the exponential is a decreasing function of $k_l$, as  its derivative is  $$(\log (2(1+\delta)\alpha_1) + 1+\log (\frac{|A_0| - k_l}{|A_0|}) < 0$$ Thus, for all $l$, it is less than or equal to its value at $k_l = \lceil\frac{\log^2 N}{N}|A_0|\rceil$. For   $k_l = \lceil\frac{\log^2 N}{N}|A_0|\rceil$, we have $(|A_0|-k_l)\log \frac{|A_0|}{|A_0| - k_l} = k_l(o(1)+1)$, and $P(M_1) \geq 1- N^2e^{k_l(\log(2(1+\delta)\alpha_1)+1+o(1))}\rightarrow 1$.

\noindent\textbf{Proof of $M_2$ happening with probability approaching one:} For the event $M_2$, we only need to check $B_{2, |A_1|}$ because $B_{2,l}\subseteq B_{2, l'}$ for all $l \leq l'$. By Lemma \ref{lem:tailDecays} part (2),  $\sup_i \sup_l \frac{P(p_{i, l} \leq \alpha_2)}{\alpha_2}\leq (1+\delta)$.
 As a result, $|B_{2, |A_1|}|$ is at most $y\sim$Bin($|A_0|$, $(1+\delta)\alpha_2$). The variable $y$ has mean $(1+\delta)\alpha_2|A_0|$ and variance bounded by $(1+\delta)\alpha_2|A_0|$. We apply Chebyshev's inequality and reach our conclusion $P(M_2) = P(|B_{1, N\gamma}| \leq 2\alpha_2|A_0|) \geq P(y \leq 2\alpha_2 |A_0|) \geq 1- \frac{(1+\delta)}{(1-\delta)^2\alpha_2|A_0|}\rightarrow 1$.
 
\noindent\textbf{Proof of $M_3$ happening with probability approaching one:} Let $\tilde{x}^2_{\alpha}$ be the upper $\alpha^{th}$ quantile of $\{x^2_i, \;i\in A_0\}$. It is sufficient to consider $A_{\alpha}$, the set of $x^2_i$ whose value is no smaller than $\tilde{x}_{\alpha}$, so $|A_{\alpha}| = \lceil \alpha |A_0|\rceil$. Let $D_1 = \{\forall l=0,1,\ldots, |A_1|,  \frac{\tilde{x}^2_{t_l}}{2(\tau^2+C)\log \frac{1}{t_l}} \leq 1\}$.  Let $z$ be a standard normal variable. For each $i\in A_0$, we have $P(x^2_i \geq 2(\tau^2+C)\log \frac{1}{t_l}) \leq P(z^2 \geq 2(\tau^2+C)\log \frac{1}{t_l})= 2\tilde{\Phi}(\sqrt{2\log \frac{1}{t_l}}) \overset{eq.(\ref{eq:Millnorm})}{\leq }\frac{t_l}{\sqrt{\pi \log \frac{1}{t_l}}}$. Because $|A_0|>cN$ for some positive constant $c$ and $t_l\in[\frac{\log^2 N}{N}, \alpha_2)$, we  apply the sterling's approximations:
\begin{align*}
P(D^c_1)\leq  N \max^{|A_1|}_{l = 0}\left(\begin{array}{c} |A_0|\\ \lceil|A_0|t_l\rceil  \end{array}\right)P(x^2_i \geq 2(\tau^2+C)\log \frac{1}{t_l})^{\lceil|A_0| t_l\rceil}\leq \max^{|A_1|}_{l=0}\frac{N(\frac{t_l}{\sqrt{\pi \log \frac{1}{t_l}}})^{|A_0| t_l}}{t_l^{|A_0| t_l}(1-t_l)^{|A_0|-|A_0| t_l}}
\end{align*}
As $t_l$ goes to 0 in $l$, for large enough $N$, we have
\[
\frac{1}{t_l^{|A_0|t_l}(1-t_l)^{|A_0|(1-t_l)}}=\exp\left(-|A_0|t_l\log t_l-|A_0|(1-t_l)\log (1-t_l)\right) \leq \exp(-|A_0|t_l\log t_l+2|A_0|t_l)\]
As $t_l \geq \frac{\log^2N}{N}$, $|A_0| > cN$, for $N$ large enough,  we have  $P(D^c_1) \leq N\exp(|A_0|t_l(2-\frac{1}{2}\log \log \frac{1}{t_l} ))\rightarrow 0$.

Now we show $M_3$ happens high probability. As $t_l \geq \frac{\log^2N}{N}$ and $|A_0| > cN$,  for $N$ large enough, we have $0.3(\tau^2+C)t_l |A_0|\log \frac{1}{t_l} \geq 8(\tau^2+C)\log N$. Let  $M_3'= \{\forall l = 0,1,\ldots, |A_1|, \; \sum_{i\in A_{t_l}} x^2_i \leq 2.2(\tau^2+C)t_l |A_0|\log \frac{1}{t_l}+8(\tau^2+C)\log N\}$, we have $M_3'\subseteq M_3$ for large $N$. Let $\tilde{A}_{t_l}$ be nulls such that  $x^2_i\geq 2(\tau^2+C)\log \frac{1}{t_l}$. When $D_1$ is true, if $x^2_i$ exceeds $2(\tau^2+C)\log \frac{1}{t_l}$, it must also exceeds $\tilde{x}^2_{t_l}$, in other words, $\tilde{A}_{t_l} \subseteq A_{t_l}$. Since $|A_{t_l}| =\lceil t_l |A_0|\rceil$ and $x^2_i-2.2(\tau^2+C)\log\frac{1}{t_l} < 0$ for $i\neq \tilde{A}_{t_l}$, for $N$ large enough:
\begin{align*}
\{M_2',\; D_1\}= &\{\forall l = 0,1,\ldots, |A_1|, \; \sum_{i\in A_{t_l}} x^2_i \leq 2.2(\tau^2+C)t_l |A_0|\log \frac{1}{t_l}+8(\tau^2+C)\log N, D_1\}\\
\subseteq &\{\forall l = 0,1,\ldots, |A_1|,\underbrace{\sum_{i\in \tilde{A}_{t_l}} (x^2_i -2.2(\tau^2+C)\log \frac{1}{t_l})}_{I_{l}}) \leq 8(\tau^2+C)\log N\} \cap D_1
\end{align*}
Let  $w_i = \frac{\tau^2+C}{\tau^2+\sigma^2_i}\in [1, C]$, $z_i = \frac{x_i}{\sqrt{\tau^2+\sigma^2_i}}\sim N(0,1)$ for $i\in A_0$. Rearrange $I_{l}$:
\begin{align*}
I_{l} &= \sum_{i\in A_0}(x^2_i -2.2(\tau^2+C) \log \frac{1}{t_l})\mathbbm{1}_{x^2_i\geq2(\tau^2+C) \log \frac{1}{t_l}} =(\tau^2+C) \sum_{i\in A_0}\frac{1}{w_i}(z^2_i -2.2w_i \log \frac{1}{t_l})\mathbbm{1}_{z^2_i\geq2w_i\log \frac{1}{t_l}}
\end{align*}
Let $y_i :=\frac{1}{w_i}(z^2_i -2.2w_i \log \frac{1}{t_l})\mathbbm{1}_{z^2_i\geq2w_i \log \frac{1}{t_l}}$. The moment generating function of $y_i$ is($\lambda< \frac{w_i}{2}$): 
\begin{align*}
M_{y_i}(\lambda) &= 2\int_{z_i \geq \sqrt{2w_i\log \frac{1}{t_l}}} \exp(\frac{1}{w_i}(z^2_i -2.2w_i\log \frac{1}{t_l} )\lambda)\frac{1}{\sqrt{2\pi}}\exp(-\frac{z^2_i}{2})d z_i +P(z^2_i \leq 2w_i\log \frac{1}{t_l}) \\
=& \frac{2}{\sqrt{1-2\frac{\lambda}{w_i}}}\exp(-2.2\lambda\log \frac{1}{t_l})\tilde{\Phi}(\sqrt{(1-2\frac{\lambda}{w_i})2w_i\log \frac{1}{t_l}})+(1-2\tilde{\Phi}(\sqrt{2w_i\log \frac{1}{t_l}}))
\end{align*}
By the Mill's ratio bound (\ref{eq:Millnorm}),  we have
\[
M_{y_i}(\lambda) \leq  1+2t_l^{w_i}(\frac{t_l^{0.2\lambda}}{(1-2\frac{\lambda}{w_i})\sqrt{4w_i\pi\log \frac{1}{t_l}}} - \frac{\sqrt{2w_i\log \frac{1}{t_l}}}{\sqrt{2\pi}(1+2w_i\log  \frac{1}{t_l})})
\]
Take $\lambda =\frac{1}{4}$. Because we have $t_l\rightarrow 0$ over $l$,  for $N$ large enough, we have $M_{y_i}(\frac{1}{4})  \leq 1$. As a result, for $N$ large enough: $P(I_{l} \geq 8(\tau^2+C)\log N) =P(\sum_{i\in A_0} y_i\geq 8\log N)  \leq \frac{\prod_{i\in A_0} (M_{y_i}(\frac{1}{4}))}{\exp(2\log N)} \leq \frac{1}{N^{2}} $, and
\begin{align*}
 P(\max^{|A_1|}_{l = 0} I_{l} \leq 8(\tau^2+C)\log N) \rightarrow 1\Rightarrow P(M_3')\rightarrow 1\Rightarrow P(M_3)\rightarrow 1
\end{align*}

\noindent\textbf{Proof of $M_4$ and $M_5$ happening with probability approaching one:} For a small constant $c$, we define $I_{1,l} = \sum_{i\in A_1}(x^2_i - \hat{\sigma}^2_i - \tau^2_l)\mathbbm{1}_{\hat{\sigma}^2_i \geq c}\mathbbm{1}_{x^2_i\leq \tilde{F}^{-1}_i(\alpha_2)(\tau^2_l+\hat{\sigma}^2_i)}$ and $I_{2,l} = \sum_{i\in A_1}(x^2_i - \hat{\sigma}^2_i - \tau^2_l)\mathbbm{1}_{\hat{\sigma}^2_i < c}\mathbbm{1}_{x^2_i\leq \tilde{F}^{-1}_i(\alpha_2)(\tau^2_l+\hat{\sigma}^2_i)}$. We want to show $I_{1,l}+I_{2,l}\geq -(\tau^2+1)(1-\gamma)\sqrt{|A_1|\log N}$ for all $l$ with high probablity. Let $\epsilon$ and $L$ be the constants in Assumption \ref{ass:ass2}.  We have $\tau^2_l\leq \tau^2$ and for the smallest $\tau^2_{|A_1|}$, we have $\frac{\tau^2-\tau^2_{|A_1|}}{\tau^2+1}\rightarrow 0$. As $\alpha_2\rightarrow 0$, for large enough $N$, we have $(\tilde{F}^{-1}_i(\alpha_2)-1-\epsilon)(\tau^2_l+1) > L(\tau^2+1)$ and 
\[
\{x^2_i - (1+\epsilon)\hat{\sigma}^2_i\leq L(\tau^2+1)\}\cap \{\hat{\sigma}^2_i\geq c\}\subseteq \{x^2_i\leq \tilde{F}^{-1}_i(\alpha_2)(\tau^2_l+\hat{\sigma}^2_i)\}\cap \{\hat{\sigma}^2_i\geq c\}
\]
If we include any point in $\{x^2_i - (1+\epsilon)\hat{\sigma}^2_i\geq L(\tau^2+1)\}$, we increase $I_{1,l}$. Using also $\tau^2\geq \tau^2_l$, we have $I_{l,1} \geq  \sum_{i\in A_1}(x^2_i - \hat{\sigma}^2_i - \tau^2)\mathbbm{1}_{\hat{\sigma}^2_i \geq c}\mathbbm{1}_{x^2_i - (1+\epsilon)\hat{\sigma}^2_i\leq L(\tau^2+1)}$.  If we include any point in $\{x^2_i - (\hat{\sigma}^2+\tau^2_l )\geq 0\}$, we increase $I_2$. Thus, we have $I_2  \geq -(c+\tau^2_l)\sum_{i\in A_1}\mathbbm{1}_{\hat{\sigma}^2_i < c}\mathbbm{1}_{x^2_i -(\hat{\sigma}^2+\tau^2_l )\leq 0}\geq  -(c+\tau^2)\sum_{i\in A_1}\mathbbm{1}_{\hat{\sigma}^2_i < c}\mathbbm{1}_{x^2_i -(1+\epsilon)\hat{\sigma}^2\leq L(\tau^2+1)}$. Hence, we have
\[
I_{1,l}+I_{2,l}\geq \sum_{i\in A_1}[(x^2_i - \hat{\sigma}^2_i - \tau^2)\mathbbm{1}_{\hat{\sigma}^2_i \geq c} - (\tau^2+c)\mathbbm{1}_{\hat{\sigma}^2_i < c}]\mathbbm{1}_{x^2_i - (1+\epsilon)\hat{\sigma}^2 \leq L(\tau^2+1)}
\]
The lower bounds no longer involve  $l$. Let $y_i =[(x^2_i - \hat{\sigma}^2_i - \tau^2)\mathbbm{1}_{\hat{\sigma}^2_i \geq c} - (\tau^2+c)\mathbbm{1}_{\hat{\sigma}^2_i < c}]\mathbbm{1}_{x^2_i - (1+\epsilon)\hat{\sigma}^2_i\leq L(\tau^2+1)}$ and let $\tilde{A} = \{i: x^2_i - (1+\epsilon)\hat{\sigma}^2_i\leq L(\tau^2+1)\}$, we have $E[y_i] = P(\tilde{A})E[(x^2_i - \hat{\sigma}^2_i - \tau^2)\mathbbm{1}_{\hat{\sigma}^2_i \geq c} - (\tau^2+c)\mathbbm{1}_{\hat{\sigma}^2_i < c}|\tilde{A}]$. For any $y\geq 0$, we have 
\begin{align*}
&E[\mathbbm{1}_{\hat{\sigma}^2_i < y}|\tilde{A}] = \int_{t\geq 0}P( \hat{\sigma}^2_i \leq y | (1+\epsilon)\hat{\sigma}^2_i\geq t - (L+1))d P(x^2_i > t) \leq  \int_{t\geq 0}P( \hat{\sigma}^2_i \leq y)d P(x^2_i > t) = P(\hat{\sigma}^2_i \leq y)\\
&E[\hat{\sigma}^2_i|\tilde{A}] = \int_{y\geq 0} y d P(\hat{\sigma}^2_i > y|\tilde{A})\geq  \int_{y\geq 0} y d P(\hat{\sigma}^2_i > y) = E[\hat{\sigma}^2_i] \\
&E[x^2_i\mathbbm{1}_{\hat{\sigma}^2_i <c}|\tilde{A}] = P(\hat{\sigma}^2_i <c)\int^\infty_{t=0}E[x^2_i|x^2_i \leq (1+\epsilon)t+L(\tau^2+1)]\frac{dP(\hat{\sigma}^2_i < t|\hat{\sigma}^2_i \leq c)}{dP(\hat{\sigma}^2_i < t)} dP(\hat{\sigma}^2_i < t)
\end{align*}
Because $E[x^2_i|x^2_i \leq (1+\epsilon)t+L(\tau^2+1)]$ is non-decreasing in $t$ and $\frac{dP(\hat{\sigma}^2_i < t|\hat{\sigma}^2_i \leq c)}{dP(\hat{\sigma}^2_i < t)}$ is non-increasing in $t$, we have
\[
E[x^2_i\mathbbm{1}_{\hat{\sigma}^2_i <c}|\tilde{A}]\leq  P(\hat{\sigma}^2_i <c)\int^\infty_{t=0}E[x^2_i|x^2_i \leq (1+\epsilon)t+L(\tau^2+1)]dP(\hat{\sigma}^2_i < t) =P(\hat{\sigma}^2_i <c)E[x^2_i|\tilde{A}] 
\]
We can now lower bound $E[(x^2_i - \hat{\sigma}^2_i - \tau^2)\mathbbm{1}_{\hat{\sigma}^2_i \geq c}|\tilde{A}] $:
\begin{align*}
E[(x^2_i - \hat{\sigma}^2_i - \tau^2)\mathbbm{1}_{\hat{\sigma}^2_i \geq c}|\tilde{A}] &\geq E[x^2_i -\hat{\sigma}^2_i|\tilde{A}] - E[x^2_i\mathbbm{1}_{\hat{\sigma}^2_i <c}|\tilde{A}]-\tau^2\geq  E[x^2_i - \hat{\sigma}^2_i|\tilde{A}]-P(\hat{\sigma}^2_i < c)E[x^2_i|\tilde{A}] -\tau^2
\end{align*}
By Assumption \ref{ass:ass2}, we have $E[x^2_i - (1+\epsilon)\hat{\sigma}^2_i|\tilde{A}] \geq (1+\epsilon)\tau^2$, if we take $c$ small enough such that $\max_{i\in A_1}P(\hat{\sigma}^2_i <c)  \leq \frac{\epsilon}{2(1+\epsilon)}$, we have
\begin{align*}
  E[y_i] &\geq E[x^2_i - \hat{\sigma}^2_i|\tilde{A}]-P(\hat{\sigma}^2_i < c)E[x^2_i|\tilde{A}] -\tau^2\\
& \geq (1-\frac{\epsilon}{2(1+\epsilon)})E[x^2_i - (1+\epsilon)\hat{\sigma}^2_i|\tilde{A}]+\frac{\epsilon}{2}E[\hat{\sigma}^2_i|\tilde{A}] - \tau^2\\
&  \geq (1-\frac{\epsilon}{2(1+\epsilon)})(1+\epsilon)\tau^2-\tau^2+\frac{\epsilon}{2} > 0
\end{align*}
We can also bound $E[y^2_i]$:
\begin{align*}
E[y^2_i] = E[(x^2_i - \hat{\sigma}^2_i - \tau^2)^2\mathbbm{1}_{\hat{\sigma}^2_i \geq c}\mathbbm{1}_{\tilde{A}} +(\tau^2+c)^2\mathbbm{1}_{\hat{\sigma}^2_i < c}\mathbbm{1}_{\tilde{A}}]
\end{align*}
When $\tilde{A}$ is true, $(x^2_i - \hat{\sigma}^2_i - \tau^2)^2 \leq \max((\hat{\sigma}^2_i+ \tau^2)^2, (L(\tau^2+1)+\epsilon \hat{\sigma}^2_i)^2)$, therefore, we have
\begin{align*}
E[y^2_i] &\leq (\tau^2+c)^2+E[(\hat{\sigma}^2_i+ \tau^2)^2]+E[(L(\tau^2+1)+\epsilon \hat{\sigma}^2_i)^2]\\
&\leq (\tau^2+1)^2+(1+\epsilon^2)E[\hat{\sigma}^4_i]+(1+L^2)(\tau^2+1)^2+2(\tau^2+L(\tau^2+1))E[\hat{\sigma}^2_i]\\
& \leq (1+L^2+(1+\epsilon^2)C^2(1+\frac{1}{m})+2(L+2)C)(\tau^2+1)^2
\end{align*}
We apply the Chebyshev's inequality to $\sum_{i\in A_1} y_i$:
\[
P(\sum_{i\in A_1}y_i \geq -(1-\gamma)\sqrt{|A_1|\log N}(\tau^2+1))\rightarrow 1
\]
As we have $I_{1,l}+I_{2,l} \leq \sum_{i\in A_1} y_i$ holds for all $l$,  hence,  $P(M_4)\rightarrow 1$. For the event $M_5$, we have $\sum_{i\in A_1}(x^2_i - \hat{\sigma}^2_i-\tau^2) \leq \sum_{i\in A_1}(x^2_i - \hat{\sigma}^2_i-\tau^2)\mathbbm{1}_{\tilde{A}}\leq \sum_{i\in A_1}y_i$. Therefore $P(M_5)\rightarrow 1$.

\section{Estimation procedures}
\label{app:procedures}
In this section, we give the details of the truncated MLE estimate and the CM estimate of the spreading factor $\tau^2$.

\noindent\textbf{Truncated $MLE$}: Let $C$ be a normalization constant depending on the context, the likelihood function of the observed points from null distribution with mean level $\mu_i$ in terms of sufficient statistics $\bar{x}_i$ and $\hat{\sigma}^2_i$($\hat{\sigma}^2_i = m\hat{\sigma}^2_{\bar{x}_i}$) is
\[
f = C\prod_{i\in A_0}(\tau^2)^{-\frac{1}{2}}\sigma_i^{-(m-k)} e^{-\frac{(\bar{x}_i - \mu_i)^2}{2\tau^2}}e^{-\frac{(m-k)\hat{\sigma}^2_i}{2\sigma^2_i}}
\]
marginalized  out $\mu_i$:
\begin{equation}
f = C\prod_{i\in A_0}(\tau^2+\sigma^2_{\bar{x}_i})^{-\frac{1}{2}}\sigma_i^{-(m-k)} \exp(-\frac{(m-k)\hat{\sigma}^2_i}{2\sigma^2_i}-\frac{\bar{x}^2_i}{2(\tau^2+\hat{\sigma}^2_{\bar{x}_i})})
\end{equation}
For points in $A_0$ with mean difference $\bar{x}_i$ in $(-\delta_0, \delta_0)$,  for a positive value $\delta_0$, this truncated likelihood function is:
\[
f_{truncated} = C\prod_{i\in A_0, \bar{x}_i \in (-\delta_0, \delta_0)}\frac{I_{[ \bar{x}_i \in (-\delta_0, \delta_0)]}}{H(\tau^2, \sigma^2_{\bar{x}_i})}(\tau^2+\sigma^2_{\bar{x}_i})^{-\frac{1}{2}}\sigma_i^{-(m-k)} \exp(-\frac{(m-k)\hat{\sigma}^2_i}{2\sigma^2_i}-\frac{\bar{x}^2_i}{2(\tau^2+\hat{\sigma}^2_{\bar{x}_i})})\]
where
\[
H(\tau, \sigma^2_{\bar{x}_i}) = \int_{x \in [-\delta_0, \delta_0]} \frac{1}{\sqrt{2\pi(\tau^2+\sigma^2_{\bar{x}_i})}}\exp(-\frac{x^2}{2(\tau^2+\sigma^2_{\bar{x}_i})})
\]
Assuming that  the observed $\{\bar{x}_i, \forall i\in A_1\}$ will not fall into the range  $(-\delta_0, \delta_0)$, we have
\[
f_{truncated} = C\prod_{\bar{x}_i \in (-\delta_0, \delta_0)}\frac{I_{[ \bar{x}_i \in (-\delta_0, \delta_0)]}}{H(\tau^2, \sigma^2_{\bar{x}_i})}(\tau^2+\sigma^2_{\bar{x}_i})^{-\frac{1}{2}}\sigma_i^{-(m-k)} \exp(-\frac{(m-k)\hat{\sigma}^2_i}{2\sigma^2_i}-\frac{\bar{x}^2_i}{2(\tau^2+\hat{\sigma}^2_{\bar{x}_i})})
\]
\begin{align}
&l_{truncated}= -\log f_{truncated} \nonumber\\
&= C+\sum^N_{i=1}I_{[\bar{x}_i \in (-\delta_0, \delta_0) ]}(\log H_i + \frac{\log (\tau^2+\sigma^2_{\bar{x}_i})}{2} + \frac{m-k}{2}\log \sigma^2_i +\frac{(m-k)\hat{\sigma}^2_i}{2\sigma^2_i}+ \frac{\bar{x}^2_i}{2(\tau^2+\sigma^2_{\bar{x}_i})})
\end{align}
We can find the minimizer to the above target function by iteratively updating $\tau^2$ and $\{\sigma^2_i:  -\delta_0 \leq \bar{x}_i  \leq \delta_0\}$. We start from $\tau^2 = 0$ and do the following,
\begin{align*}
&\text{For } \tau^2 \text{ fixed, solutions to }\{\sigma^2_i\}:\\
& \hat{\sigma}^2_i = \arg\min_{\sigma^2_i} \log H_i+\frac{\log (\tau^2+\sigma^2_{\bar{x}_i})}{2} + \frac{m-k}{2}\log \sigma^2_i +\frac{(m-k)\hat{\sigma}^2_i}{2\sigma^2_i}+ \frac{\bar{x}^2_i}{2(\tau^2+\sigma^2_{\bar{x}_i})}\\
&\text{For }  \{\sigma^2_i\} \text{ fixed, solution to }\tau^2:\\
& \hat{\tau}^2_i = \arg\min_{\tau^2} \sum^N_{i=1}I_{[\bar{x}_i \in (-\delta_0, \delta_0) ]}(\log H_i + \frac{\log (\tau^2+\sigma^2_{\bar{x}_i})}{2} +\frac{\bar{x}^2_i}{2(\tau^2+\sigma^2_{\bar{x}_i})})
\end{align*}

\noindent\textbf{CM}: The marginal density of $\bar{x}_i$(marginalized over the index $i$) for all genes can be written as following
\begin{equation}
f(x|\tau) \sim \frac{1}{N}\sum_{i\in A_0}\frac{e^{-\frac{x^2}{2(\sigma^2_{\bar{x}_i}+\tau^2)}}}{\sqrt{2\pi (\sigma^2_{\bar{x}_i}+\tau^2)}} +\frac{1}{N}\sum_{i\in A_1} h_i(x)
\end{equation}
where $h_i(.)$ is the density function for $\bar{x}_i$ when  $i\in A_1$, which is $g_i(\mu)$ convolved with the a normal distribution describing the noise in $\bar{x}_i$: $h_i(x) = \int g_i(\mu)\frac{1}{\sqrt{2\pi \sigma^2_{\bar{x}_i}}}e^{-\frac{(x-\mu)^2}{2\sigma^2_{\bar{x}_i}}} d\mu$.

Like in truncated MLE method, we assume that $A_1$'s contribution to the region $[-\delta_0, \delta_0]$ is negeligiable. Doing a first order Taylor expansion of the marginal density function $f(x|\tau)$, we have
\begin{align*}
f(x|\tau) &\approx \frac{1}{N} \sum^N_{i=1} \frac{1}{\sqrt{2\pi(\sigma^2_{\bar{x}_i}+\tau^2)}}(1-\frac{x^2}{2(\sigma^2_{\bar{x}_i}+\tau^2)})\\
l(x) = -\log f(x) & \approx C +\sum^N_{i=1}\frac{x^2}{(\sigma^2_{\bar{x}_i}+\tau^2)^{\frac{3}{2}}\sum_{i \in A_0}\frac{1}{\sqrt{\sigma^2_{\bar{x}_i}+\tau^2}}} 
\end{align*}

We find $\tau^2$ by fitting a polynomial function to $(x,\hat{l}(x))$, where $\hat{l}(x)$ is the negative log transformation of the empirical marginal density function, estimated by binning our observations in $[-\delta_0, \delta_0]$--such binning and fitting steps are also used by the R function  \textbf{locfdr}. As a result, we  can get the estimated $\tau^2$ simply as following:

\begin{enumerate}
\item Use the R function \textbf{locfdr} with central  matching approach to find the coefficients of the  second term in the regression fit
\[
\hat{l}(x) = \beta_0 + \beta_1x+\beta_2 x + ....
\]
and denote it as $\hat{\beta}_2$.
\item Use the relationships below to fo a grid search of $\tau^2$:
\begin{align*}
&\sum^N_{i=1}\frac{1}{2(\sigma^2_{\bar{x}_i} + \tau^2)^{\frac{3}{2}}\sum^N_{i=1 }\frac{1}{\sqrt{\sigma^2_{\bar{x}_i} + \tau^2}}} = \hat{\beta_2}
\end{align*}
with $\sigma^2_{\bar{x}_i}$ replaced by  $\hat{\sigma}^2_{\bar{x}_i}$.
\end{enumerate}

Intuitively, the three procedures are different in several perspectives:
\begin{enumerate}
\item ITEB starts by treating the full data set as null and iteratively removing genes with large values,  it usually ends up estimating using a set of genes much larger than the other two methods.  As a result, it is be able to utilize more information from the data, but it suffers more from initially overestimating $\tau^2$ when $\gamma$ is large(We care about small $\gamma$ in our case).
\item CM relies on the first order Taylor expansion of the log likelihood  around a small region near $0$, as well as the plug-in variance estimates, and it also needs to plug in variance estimates in the denominator, which makes it unstable.
\item Both the truncated MLE method and CM need to know the specific form of the likelihood for the null distribution, while ITEB uses only the moments, which makes its application to complicated distributions straightforward. Also, both CM and truncated MLE have removed a significant proportion of data from the beginning, and as a result, their estimates can have larger variance compared with that of ITEB when $\gamma$ is small.
\end{enumerate}
In Appendix \ref{app:sim}, we compare  performances  for the three estimates in different scenarios and  discuss their strengths and weaknesses. 

\section{Simulation: Estimate of $\tau$}
\label{app:sim}
For simplicity, we focus on the one-sample setting and and generate data under various values of $\tau$ and non-null proportion $\gamma= \frac{|A_1|}{N}$. Specifically, we  fix $N = 15000, m = 10$, which is of the same order  as typical knock-down data. For any given $\tau$ and $\gamma$, where $p = 0,1\%,...,10\%$ and $\tau = 0, 0.1,...,1, 1.5, 2, 2.5,3$, we generate the data as below.
\begin{enumerate}
\item Generate $\mu_i$s:
 $\mu_i \sim  \left\{\begin{array}{ll}N(0, \tau^2)&\forall i\in A_0\\
\pm U[1, \max(3, 10\tau)]&\forall i\in A_1\\
\end{array} \right.$

 where $U[1,\max(3, 10\tau)]$ is the uniform distribution between $1$ and $\max(3, 10\tau)$, and the signs of $\mu_i$s will be half positive and half negative.

\item Generate variances for genes in one of the two settings:
\begin{enumerate}
\item Independently generate $\sigma^2_i \sim \chi^2_1$.
\item Sample $\sigma^2_i$ from its empirical distribution in the real data set, scaled to have mean level 1.
\end{enumerate}
\end{enumerate}
We used the three approaches to estimate $\tau$ with  $(\alpha_1, \alpha_2) = (0.1, 0.01)$ for ITEB and leave out proportion to be $0.2$ both for truncated MLE and CM. We repeat the simulations 20 times and plot the relative mean relative errors  $\frac{\sum^{20}_{i=1}|\hat{\tau}^2-\tau^2|}{20(\tau^2+0.1)}$ in  Figure \ref{fig:simulationTau_Gaussian}. We have also consider the case where the data is not normal by generating parameters in the same way but with Laplacian distributed noise(means and variances are matched), results are given in  Figure \ref{fig:simulationTau_Lap}.

From the simulation results, we can see that (1)CM does not seem to be a good approach estimating $\tau$. (2) ITEB's performance is as good as Truncated MLE, if not better, across the parameters we have considered. 

While the performances for both ITEB and truncate MLE are reasonably good, there is a huge time difference. In our simulation, the truncated MLE's run time is more than 100 times of that of ITEB: the average run time per round for the ITEB  is  about $0.1$ second, while it is $13$ seconds for the the truncated MLE.

Overall, we consider ITEB to be a good substitute for the truncated MLE estimate because of its simplicity and good performance, and both methods are preferable over CM.
\begin{figure}[h]
\begin{center}
\includegraphics[width = .49\textwidth, height = .6\textwidth]{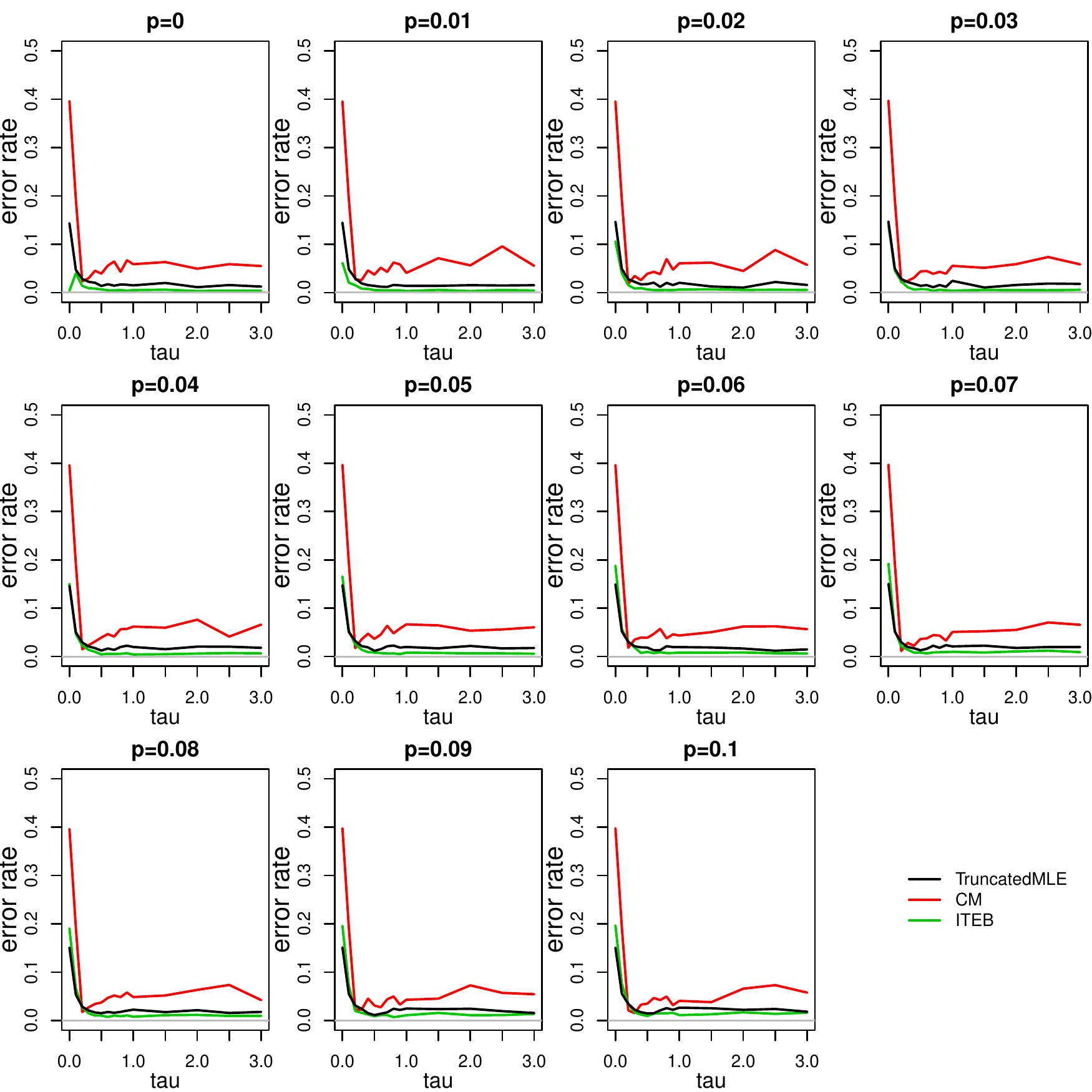}
\includegraphics[width = .49\textwidth, height = .6\textwidth]{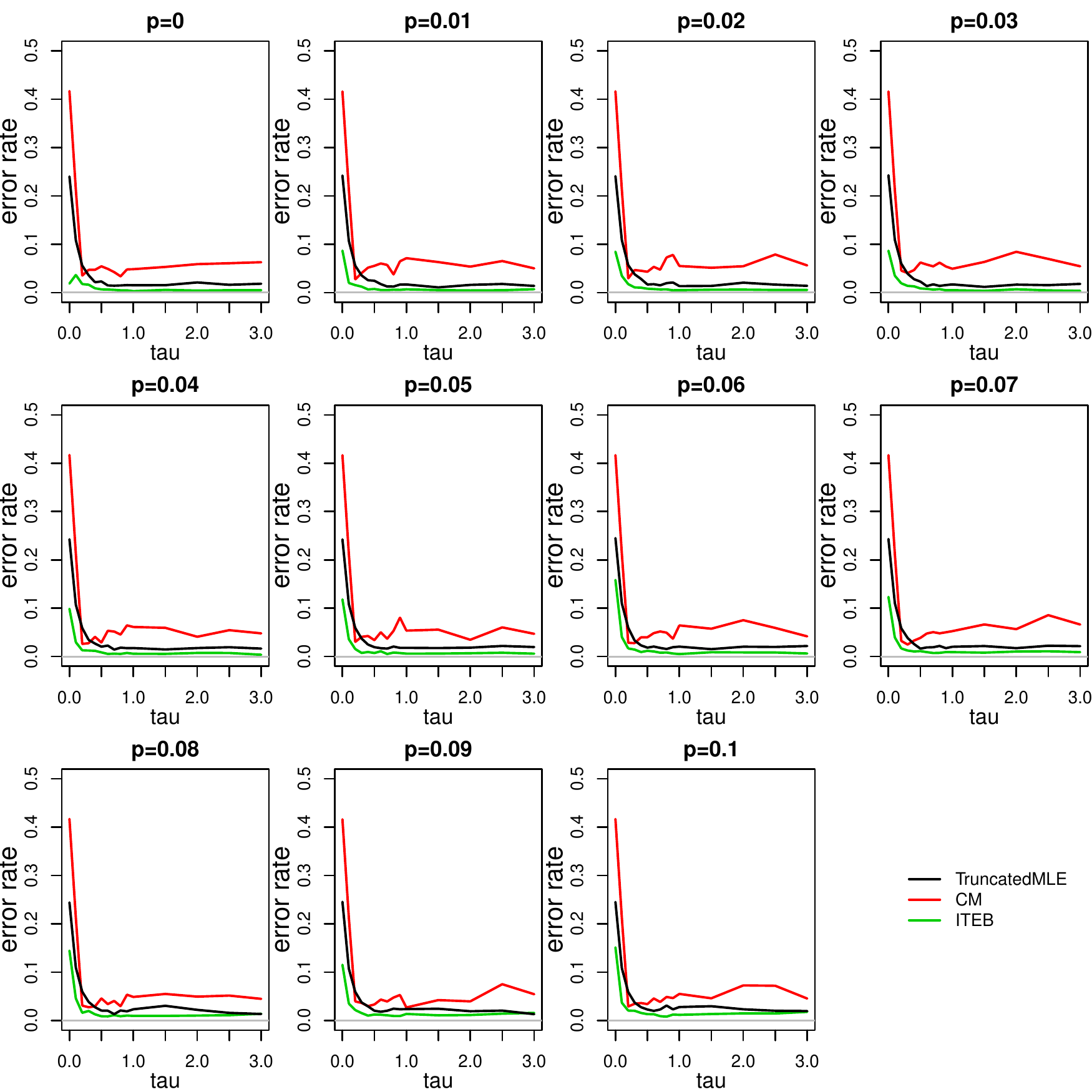}
\end{center}
\caption{\em Gaussian noise, setting (a) and (b): The left half plot is the plot of the estimated relative error versus $\tau$ from setting (a)and the right half plot is that from setting (b)}\label{fig:simulationTau_Gaussian}
\end{figure}
\begin{figure}[h]
\begin{center}
\includegraphics[width = .49\textwidth, height = .6\textwidth]{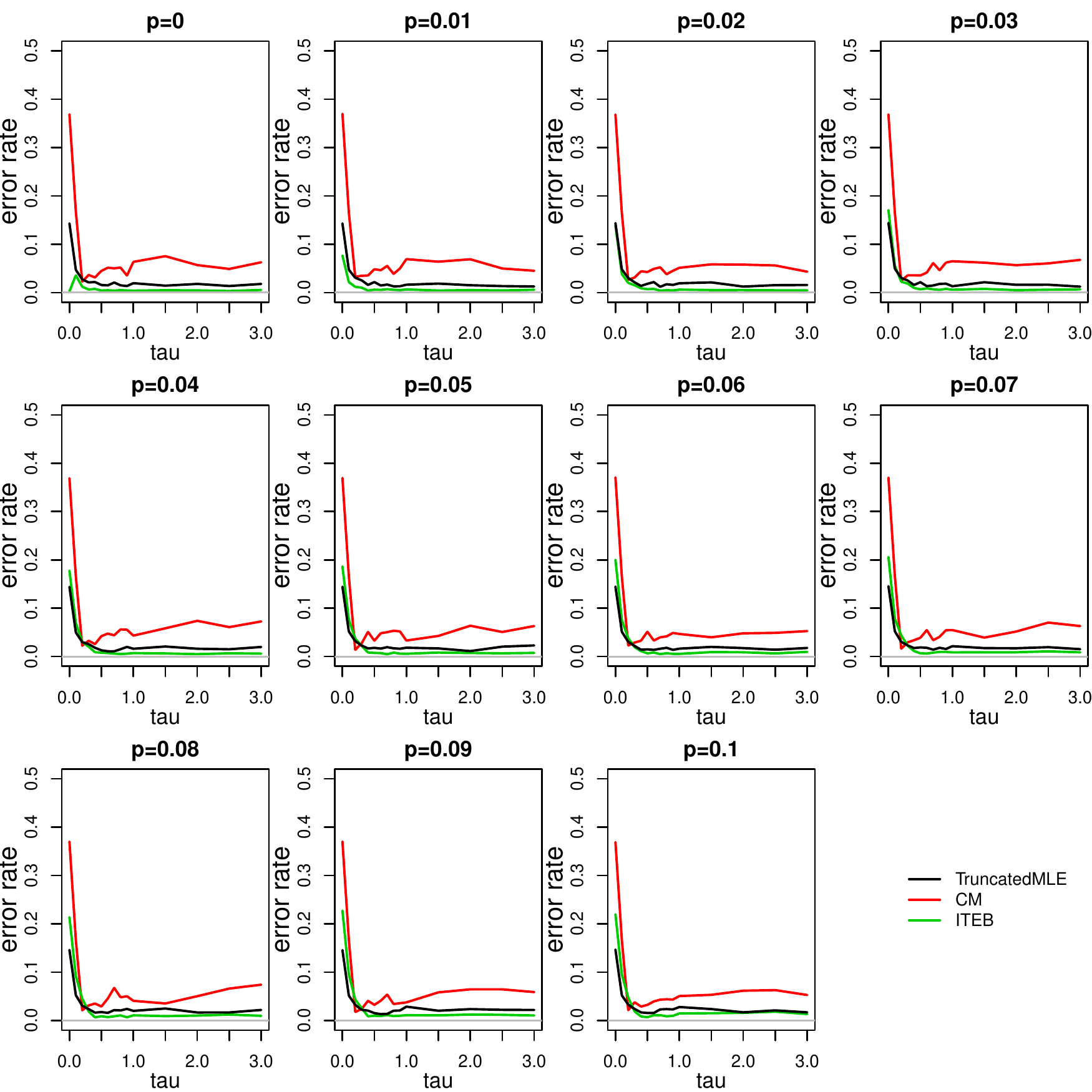}
\includegraphics[width = .49\textwidth, height = .6\textwidth]{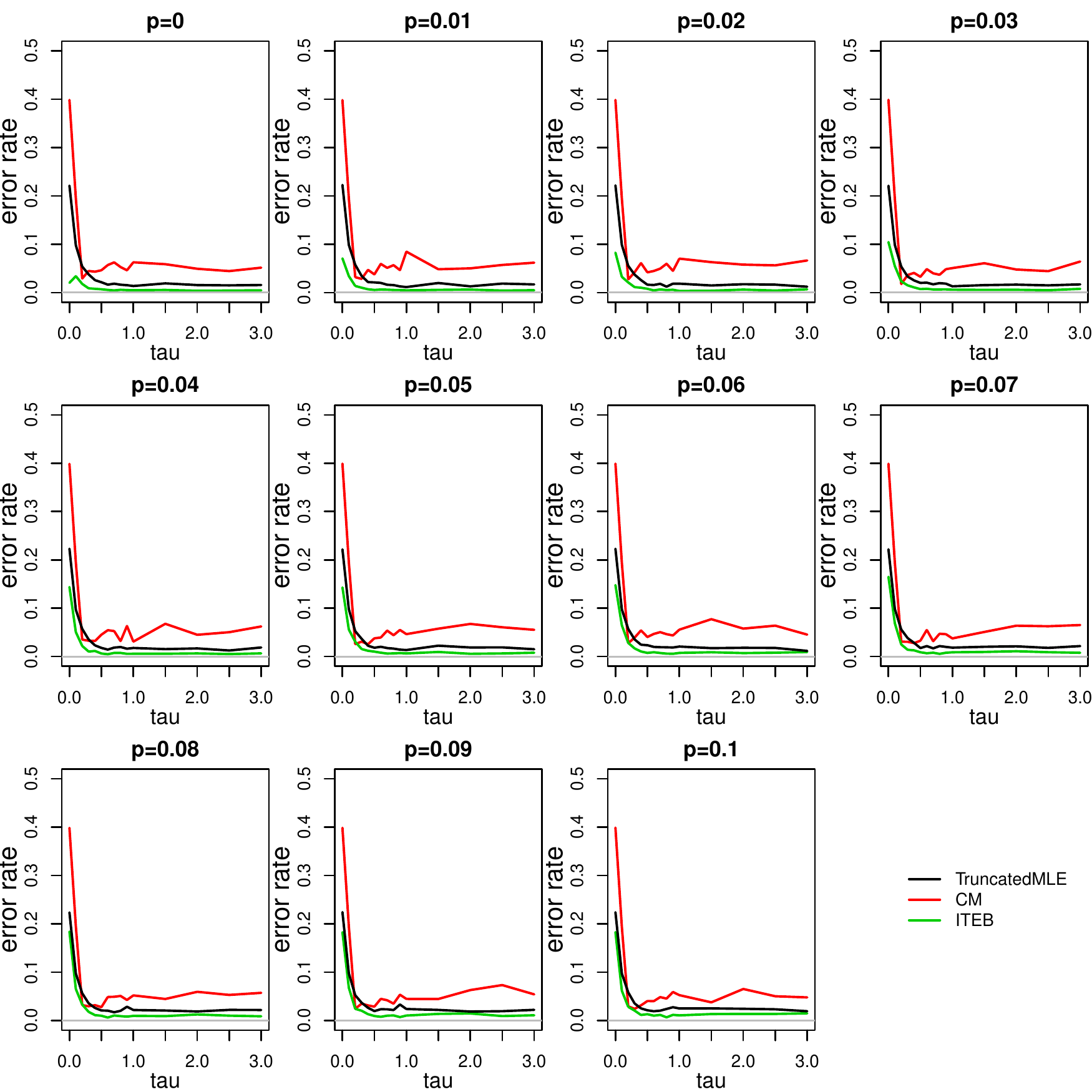}
\end{center}
\caption{\em Laplacian noise, setting (a) and (b): The left half plot is the plot of the estimated relative error versus $\tau$ from setting (a)and the right half plot is that from setting (b)}\label{fig:simulationTau_Lap}
\end{figure}
\section{Materials  and Methods for the knock-down experiment}
 \label{app:experiment}
 In this section, we provide more details about how the data is generated and justification of pooling data across days.
 
\noindent\textbf{Cell Culture:} Mouse ES cell line R1 was obtained from Dr. Douglas Melton lab (Harvard University, MA) and cultured under standard conditions. The cells were maintained on gelatin-coated dishes in RPMI knockout medium with 15\% knockout serum replacement (KSR), 2 mM L-glutamine, 1 mM non-essential amino acids, 0.55 mM 2-mercaptoethanol (Invitrogen, CA), and 1000 units/mL murine leukaemia inhibitory factor (Chemicon International, CA). Cells were incubated in a 5\% CO2--air mixture at $37^{\rm o}$C. Cultures were routinely passaged with 0.25\% trypsin-EDTA (Invitrogen, CA) and split 1:8 every 2 days. Normal karyotype of ESC was routinely confirmed by analysis of chromosome spreads.

\noindent\textbf{RNA Interference:}  RNA interference (RNAi) experiments were performed with Nucleofector technology. Briefly, 12 $\mu$d of plasmid DNA was transfected into $3.5\times10^6$ mouse ES cells using the Mouse ES cell Nucleofector kit (Lonza, Switzerland).  After nucleofection, the cells were incubated in $500\;\mu$l warm ES medium for 15 min. Then, the cells were split into four gelatin-coated 60-mm tissue culture plates containing 5 ml of warm ES medium. Puromycin selection was introduced 18 h later at 1 $\mu$ g/ml, and the medium was changed daily. 30 h, 48 h, and 72 h after puromycin selection, the cells were collected for RNA isolation.

\noindent\textbf{Microarray and Data Processing:} Microarray hybridizations were performed on the MouseRef-8 v2.0 expression beadchip arrays (Illumina, CA). To prepare sample, 200 ng of total RNA was reverse transcribed, followed by a T7 RNA polymerase-based linear amplification using the Illumina TotalPrep RNA Amplification kit (Applied Biosystems, CA). After amplification, 750 ng of biotin-labeled cRNA was hybridized to gene specific probes attached to the beads, and the expression levels of transcripts were measured simultaneously. 

\bibliographystyle{agsm}
\bibliography{leying}
\end{document}